\documentclass[epj]{svjour}
\usepackage{graphics}
\RequirePackage{graphicx}
\RequirePackage{multirow}
\RequirePackage{yhmath}
\RequirePackage{hyperref}
\usepackage{color}
\usepackage{cases}
\usepackage{subfigure}
\usepackage{dcolumn,booktabs,bm}
\usepackage{slashed}
\usepackage{amsfonts,amssymb,stmaryrd,latexsym,amsmath}
\usepackage{textcomp}
\usepackage{braket}
\usepackage{epstopdf}

\renewcommand\tabcolsep{1.5 pt}

\def\m{\mathcal}
\def\n{\nonumber}
\def\f{\frac}

\def\S{\Sigma}

\def\o{\omega}
\def\O{\Omega}
\def\L{\Lambda}

\begin{document}
\title{Hadronic Molecular States Composed of Spin-$3\over 2$ Singly Charmed Baryons}
\author{Bin Yang\inst{1}\thanks{e-mail: bin{\_}yang@pku.edu.cn } \and Lu Meng\inst{1}\thanks{e-mail: lmeng@pku.edu.cn} \and Shi-Lin Zhu\inst{1,2}%
\thanks{e-mail: zhusl@pku.edu.cn}
}                     % Do not remove

\institute{School of Physics and State Key Laboratory of Nuclear
Physics and Technology, Peking University, Beijing 100871, China\label{addr1} \and  Collaborative Innovation Center of Quantum Matter, Beijing 100871, China \label{addr2}}
\date{Received: date / Revised version: date}
\abstract{
We investigate the possible deuteron-like molecules composed of a
pair of charmed spin-$\frac{3}{2}$ baryons, or one charmed baryon
and one charmed antibaryon within the one-boson-exchange (OBE)
model. For the spin singlet and triplet systems, we consider the
couple channel effect between systems with different orbital angular
momentum. Most of the systems have binding solutions. The couple
channel effect plays a significant role in the formation of some
loosely bound states. The possible molecular states of
$\Omega_c^*\Omega_c^*$ might be
stable once produced.
%
%\PACS{
%      {PACS-key}{discribing text of that key}   \and
%      {PACS-key}{discribing text of that key}
%     } % end of PACS codes
} %end of abstract
\maketitle
%
%%%%%%%%%%%%%%%%%%%%
\section{INTRODUCTION}\label{sec_intro}
%%%%%%%%%%%%%%%%%%%%

Since the charmonium-like state $X(3872)$ was reported by the Belle
Collaboration in 2003~\cite{Choi:2003ue}, exotic states attracted
great interest around the world. Many experiment collaborations
such as BaBar, BESIII, Belle, LHCb, CDF, D0, reported discoveries of
new charmonium-like and bottomonium-like states such as
$Y(4260)$~\cite{Aubert:2005rm},
$Z_c(3900)$~\cite{Liu:2013dau,Ablikim:2013mio}, $Z_b(10610)$ and
$Z_b(10650)$~\cite{Belle:2011aa}. In 2015, LHCb reported two
hidden-charm pentaquark states $P_c(4380)$ and
$P_c(4450)$~\cite{Aaij:2015tga}. One can find the experimental and
theoretical progress about these exotic states in the recent
reviews~\cite{Chen:2016qju,Esposito:2016noz,Chen:2016spr,Lebed:2016hpi,Guo:2017jvc,Olsen:2017bmm}

It is difficult to interpret some of these states with the
conventional quark model. They may well be multi-quark states rather
than traditional $q\bar{q}$ and $qqq$ hadrons. Some of them are well studied as dynamically generated bound states or resonances~\cite{Kaiser:1995cy,Oller:2000fj,Khemchandani:2011et,Lu:2014ina,Montana:2017kjw,Liang:2017ejq}.
For the exotic states
near the threshold of two heavy hadrons, it is natural to consider
them as candidates of molecular states.
A hadronic molecular state is a loosely bound state composed of two
color-singlet hadrons. The interaction is the residual force of the
color interaction, which is usually described as one-boson-exchange
(OBE) potential. The OBE model is very successful to explain the
deuteron, a well-established hadronic molecular state composed of a
neutron and a proton. The meson exchange force together with the S-D
mixing effect render the deuteron a loosely bound state. The
binding energy is about 2.225 MeV and root-mean-square radius is
about 2.0 fm.

Voloshin and Okun proposed the hadronic molecular
composed of two charmed mesons about forty years
ago~\cite{Voloshin:1976ap}. De Rujula et al. also used the molecular
model to interpret the $\psi(4040)$ as a $D^*\bar{D}^*$
molecule~\cite{DeRujula:1976zlg}. T\"ornqvist used the
one-pion-exchange (OPE) potential to calculate the possible
molecular state composed of one charmed meson and one charmed
antimeson\cite{Tornqvist:1993ng,Tornqvist:1993vu}.

There are also many other analyses about hadronic molecular states, such as the
combination of two
mesons\cite{Wong:2003xk,Liu:2009ei,Close:2009ag,Ding:2009vj,Sun:2011uh,Li:2012ss,Zhao:2014gqa},
or two
baryons\cite{Lee:2011rka,Meguro:2011nr,Li:2012bt,Meng:2017fwb,Meng:2017udf,Vijande:2016nzk,Carames:2015sya}.
Similarly, the hidden-charm(bottom) pentaquark states can be explained as a molecular
state formed by one heavy meson and one heavy
baryon\cite{Wu:2010jy,Yang:2011wz,Chen:2015loa,Roca:2015dva,Yamaguchi:2017zmn,Shimizu:2017xrg,Yamaguchi:2016ote}.
In addition, some near threshold states might be treated as a compact core plus a molecular component, like X(3872)\cite{Ferretti:2013faa,Ferretti:2014xqa,Ferretti:2018tco}.

In the Ref.~\cite{Li:2012bt}, Li et al. calculated the possible
molecular states composed of two spin-$\frac{1}{2}$ heavy baryons
with the OPE and OBE potential, respectively. They analysed the
$\Lambda_c\Lambda_c$ system and considered the couple-channel effect
of $\Sigma_c$ and $\Sigma_c^*$. In this work, we extend the same
formalism to investigate the possible hadronic molecular states
composed of two spin-$\frac{3}{2}$ singly charmed baryons. We adopt
the OBE potential and take the couple channel effect between systems
with different orbital angular momentum into consideration.

This work is organized as follows. After the introduction, we
present the formalism in Section \ref{sec_form}, in which we
introduce the Lagrangians, coupling constants and the effective
interaction potentials. In Section \ref{sec_num} we show our
numerical results of the two heavy baryon systems. Then we discuss our results and conclude in
Section \ref{sec_dis}. We collect some useful formulae and functions
in Appendixes~\ref{app_func} and \ref{app_matrix}. We also calculate the systems composed of one heavy baryon and one heavy antibaryon. The numerical results are collected in Appendix~\ref{app_BBbarsc}.

%%%%%%%%%%%%%%%%%%%%%%%%%%%%%%%%%%%%%%%%%%%%
\section{FORMALISM}\label{sec_form}
%%%%%%%%%%%%%%%%%%%%%%%%%%%%%%%%%%%%%%%%%%%%

%%%%%%%%%%%%%%%%%%%%%%%%%
\subsection{The Lagrangian}\label{subsec_lag}
%%%%%%%%%%%%%%%%%%%%%%%%%%%%%%%%

The singly charmed baryon is composed of one charm quark and two
light quarks, which is usually treated as a diquark. In the heavy
quark limit, we can classify the singly charmed baryons with
symmetry of the light diquark. The wave function of the diquark is
as follows,
\begin{equation}
  \Psi^{total}_{qq}=\Psi^{flavor}_{qq}\otimes\Psi^{spin}_{qq}\otimes\Psi^{color}_{qq}\otimes\Psi^{spatial}_{qq}.
\end{equation}
The total wave function $\Psi^{total}_{qq}$ is antisymmetric for a
fermion system as required by Pauli Principle. The color wave
function $\Psi^{color}_{qq}$ must be in antisymmetric
$\bar{3}_c$-representation, and the spatial wave function
$\Psi^{spatial}_{qq}$ is symmetric for the ground state. As a
result, the flavor wave function $\Psi^{flavor}_{qq}$ and the spin
wave function $\Psi^{spin}_{qq}$ are correlated with each other.
When $\Psi^{flavor}_{qq}$ is symmetric in the $6_f$-representation,
$\Psi^{spin}_{qq}$ must be symmetric, which means the spin of the
diquark is 1. On the other hand, $\Psi^{flavor}_{qq}$ can also be
antisymmetric in the $\bar 3_f$-representation, and
$\Psi^{spin}_{qq}$ must be antisymmetric, i.e., the spin of diquark
is 0. Taking the spin of heavy quark into account, it is convenient
to describe the charmed baryon with its total spin and the flavor
representation of diquark. For the $6_f$-representation one, the
total spin can be $\frac{1}{2}$ and $\frac{3}{2}$. For the $\bar
3_f$-representation one, the total spin is $\frac{1}{2}$.

We denote the charmed baryons as~\cite{Yan:1992gz}:
\begin{small}
\begin{equation}
\begin{split}
B_6 &= \left[
\begin{array}{ccc}
      \S_c^{++}          & \f{1}{\sqrt{2}}\S_c^+    & \f{1}{\sqrt{2}}\Xi_c^{'+} \\
\f{1}{\sqrt{2}}\S_c^+   & \S_c^{0}                & \f{1}{\sqrt{2}}\Xi_c^{'0} \\
\f{1}{\sqrt{2}}\Xi_c^{'+}& \f{1}{\sqrt{2}}\Xi_c^{'0}&\O_c^0
\end{array}
\right],  B_{\bar{3}} = \left[
\begin{array}{ccc}
     0                 & \L_c^+              &\Xi_c^{+} \\
   -\L_c^+             &      0              &\Xi_c^{0} \\
-\Xi_c^{+}             &-\Xi_c^{0}           & 0
\end{array}
\right],
\\
B_6^* &= \left[
\begin{array}{ccc}
      \S_c^{*++}          & \f{1}{\sqrt{2}}\S_c^{*+}    & \f{1}{\sqrt{2}}\Xi_c^{*+} \\
\f{1}{\sqrt{2}}\S_c^{*+}   & \S_c^{*0}                & \f{1}{\sqrt{2}}\Xi_c^{*0} \\
\f{1}{\sqrt{2}}\Xi_c^{*+}& \f{1}{\sqrt{2}}\Xi_c^{*0}&\O_c^{*0}
\end{array}
\right].\label{heavy:baryons}
\end{split}
\end{equation}
\end{small}
We use the superscript `` * " to label spin-$\frac{3}{2}$ baryons.
The matrices of exchanged pseudoscalar and vector bosons are as
follows,
\begin{eqnarray}
%\begin{split}
\mathcal{M} &=& \left[
\begin{array}{ccc}
\f{\pi^0}{\sqrt{2}}+\f
{\eta}{\sqrt{6}}&     \pi^+               &  K^+  \\
\pi^-                   &-\f{\pi^0}{\sqrt{2}}+\f{\eta}{\sqrt{6}}&  K^0   \\
K^-                  &\bar{K}^0                    &-\f{2}{\sqrt{6}}\eta  \\
\end{array}
\right], \n
\\
\mathcal{V}^{\mu} &=& \left[
\begin{array}{ccc}
\f{\rho^0}{\sqrt{2}}+\f{\o}{\sqrt{2}}&     \rho^{+}               &  K^{*+}  \\
\rho^-                    &-\f{\rho^0}{\sqrt{2}}+\f{\o}{\sqrt{2}} &  K^{*0}   \\
K^{*-}                  &\bar{K}^{*0}                    & \phi  \\
\end{array}
\right]^{\mu}.
%\end{split}
\end{eqnarray}

Under the SU(3)-flavor symmetry, the meson exchange Lagrangians are
constructed as~\cite{Liu:2011xc}
\begin{equation}
  \m{L}=\m{L}_{\sigma hh}+\m{L}_{phh}+\m{L}_{vhh},
\end{equation}
for the scalar meson exchange
\begin{equation}\label{L_s}
  \m{L}_{\sigma hh}=-g_{\sigma B_{6}^{*}B_{6}^{*}}Tr[\bar{B}_{6}^{*\mu}\sigma B_{6\mu}^{*}],
\end{equation}
for the pseudoscalar meson exchange
\begin{equation}\label{L_ps}
  \m{L}_{phh}=-g_{pB_{6}^{*}B_{6}^{*}}Tr[\bar{B}_{6}^{*\mu}i\gamma_{5}\m{M}B_{6\mu}^{*}],
\end{equation}
and for the vector meson exchange
\begin{eqnarray}\label{L_v}
\begin{aligned}
  \m{L}_{vhh}   =&-g_{vB_{6}^{*}B_{6}^{*}}Tr[\bar{B}_{6}^{*\mu}\gamma_{\nu}\m{V}^{\nu}B_{6\mu}^{*}]
  \\
  &-i\frac{f_{vB_{6}^{*}B_{6}^{*}}}{2m_{6^{*}}}Tr[\bar{B}_{6\mu}^{*}(\partial^{\mu}\m{V}^{\nu}-\partial^{\nu}V^{\mu})B_{6\nu}^{*}].
\end{aligned}
\end{eqnarray}
The notations $g_{\sigma B_{6}^{*}B_{6}^{*}}$,
$g_{pB_{6}^{*}B_{6}^{*}}$ and $g_{vB_{6}^{*}B_{6}^{*}}$, represent
the coupling constants. $m_6^{*}$ is the mass of the spin
$\frac{3}{2}$ heavy baryon in $6_f$-representation.

%%%%%%%%%%%%%%%%%%%%%%%%%%
\subsection{Coupling Constants}\label{subsec_coupling}
%%%%%%%%%%%%%%%%%%%%%%%%%%%%%%%%%

The coupling constants in Eqs.~(\ref{L_s}-\ref{L_v}) can be
determined with the help of the nucleon-nucleon-meson vertices.
Comparing the relevant constants for heavy baryon with those for
nucleon via the quark model, we can easily get the relationship
between them. Some details can be found in Ref~.\cite{Meng:2017fwb}.
Here we list the relationships we need in this work directly,
\begin{eqnarray}
  &&g_{\sigma B_{6}^{*}B_{6}^{*}}=\frac{2}{3}g_{\sigma NN},\label{cc_S}\\
  &&g_{pB_{6}^{*}B_{6}^{*}}=\frac{6\sqrt{2}}{5}g_{\pi NN}\frac{m_{i}+m_{f}}{2m_{N}},\\
  &&g_{vB_{6}^{*}B_{6}^{*}}=2\sqrt{2}g_{\rho NN},\\
  &&g_{vB_{6}^{*}B_{6}^{*}}+f_{vB_{6}^{*}B_{6}^{*}}=\frac{6\sqrt{2}}{5}(g_{\rho NN}+f_{\rho NN})\frac{\sqrt{m_{i}m_{f}}}{m_{N}}\label{cc_V}
\end{eqnarray}
where the $g_{\sigma NN}$, $g_{\pi NN}$, $g_{\rho NN}$ and $f_{\rho
NN}$, are the nucleon-nucleon-meson coupling constants. Their
numerical values are taken from
Refs.~\cite{Machleidt:1987hj,Riska:2000gd,Machleidt:2000ge,Cao:2010km}.
For the nucleon vertices, one can also choose coupling constants for
other mesons such as $g_{\eta NN}$ and $g_{\omega NN}$. In this
work, we select three representative numerical values as mentioned
above, after consider their stability in various models. Their
values are shown in Table~\ref{Table_mass}. $m_N$ is the nucleon
mass, $m_i$ and $m_f$ are the masses of initial state and final
state baryon respectively. Thus, the numerical values of coupling
constants for different baryon-baryon-meson vertices vary slightly.
Their numerical values can be found in Table~\ref{Table_cc}. The
masses of baryons and exchanged mesons are collected in
Table~\ref{Table_mass}.

\begin{table*}[htp]
 \centering
 \caption{The relevant hadron masses~\cite{Tanabashi:2018oca} and coupling constants for the nucleon~\cite{Machleidt:1987hj,Riska:2000gd,Machleidt:2000ge,Cao:2010km}. For the multiple hadrons, their averaged masses are used.}\label{Table_mass}
\begin{tabular}{cc|cc|cc|cc}
\hline \hline Baryons & Mass(MeV) & Mesons & Mass(MeV) & Mesons &
Mass(MeV) & Couplings & Value\tabularnewline \hline $\Sigma_{c}^{*}$
& 2518.4 & $\pi$ & 137.25 & $\omega$ & 782.65 & $g^2_{\sigma
NN}/4\pi$ & 5.69\tabularnewline $\Xi_{c}^{*}$ & 2645.9 & $\eta$ &
547.85 & $\phi$ & 1019.46 & $g^2_{\pi NN}/4\pi$ &
13.6\tabularnewline $\Omega_{c}^{*}$ & 2765.9 & $\rho$ & 775.49 &
$\sigma$ & 600 & $g^2_{\rho NN}/4\pi$ & 0.84\tabularnewline
 &  &  &  &  &  & $f_{\rho NN}/g_{\rho NN}$ & 6.1\tabularnewline
\hline \hline
\end{tabular}
\end{table*}

\begin{table}[htp]
 \centering
 \caption{The coupling constants for the spin-$3\over 2$ charmed baryons.}\label{Table_cc}
\begin{tabular}{ccccc}
\hline \hline ~~Vertex~~ & ~~$g_{\sigma B_{6}^{*}B_{6}^{*}}$~~ &
~~$g_{pB_{6}^{*}B_{6}^{*}}$~~ & ~~$g_{vB_{6}^{*}B_{6}^{*}}$~~ &
~~$f_{vB_{6}^{*}B_{6}^{*}}$~~\tabularnewline \hline
$\Sigma_{c}^{*}\Sigma_{c}^{*}$ & 5.64 & 59.50 & 9.19 &
95.80\tabularnewline $\Xi_{c}^{*}\Xi_{c}^{*}$ & 5.64 & 62.51 & 9.19
& 101.12\tabularnewline $\Omega_{c}^{*}\Omega_{c}^{*}$ & 5.64 &
65.35 & 9.19 & 106.12\tabularnewline \hline \hline
\end{tabular}
\end{table}

%%%%%%%%%%%%%%%%%%%%%%
\subsection{The Effective Interaction Potentials}\label{subsec_pot}
%%%%%%%%%%%%%%%%%%%%%%

With the Lagrangians in Eqs. (\ref{L_s}-\ref{L_v}), we can get the
interaction potentials $\m{V}(Q)$ in momentum space, which can be
expanded in terms of the heavy baryon mass. We expand the potential
up to $\mathcal{O}(\frac{1}{m_Q^2})$. Then we transform the
potential to coordinate space through Fourier transformation.
\begin{equation}
  \m{V}(r)=\frac{1}{(2\pi)^3}\int d\textbf{Q}e^{i\textbf{Q}\cdot\textbf{r}}\m{V}(\textbf{Q})\cdot
  \m{F}^2(\textbf{Q})
\end{equation}
A form factor $\m{F}(Q)$ is introduced to suppress the contribution
of high momentum transfer between baryons. Within the meson exchange
framework, it is not self-consistent to keep the very short-range
interaction, which explores the inner structure of baryons. There
are many different kinds of form factors and we choose the
traditional monopole one for convenience,
\begin{equation}
  \m{F}(\textbf{Q})=\frac{\Lambda^{2}-m_{ex}^{2}}{\Lambda^{2}-Q^{2}}=\frac{\Lambda^{2}-m_{ex}^{2}}{\lambda^{2}+\textbf{Q}^{2}}.
\end{equation}
The parameter $\Lambda$ is an adjustable cutoff for suppressing the
high momentum contribution, which is 0.8-1.5 GeV suggested by the
study of the deuteron. $m_{ex}$ and $Q$ are the mass and the four
momentum of the exchanged mesons, respectively.
$\lambda^2=\Lambda^{2}-Q_0^{2}$. The specific potentials for
exchanging different mesons are as follows,
\begin{itemize}
  \item Scalar meson exchange
    \begin{equation}\label{V_S}
      \begin{split}
        V_{C}^{s}(r,\sigma)=&-C_{\sigma}^{s}\frac{g_{s1}g_{s2}}{4\pi}u_{\sigma}\bigg[H_{0}-\frac{u^{2}_{\sigma}}{8m_{A}m_{B}}H_{1}\bigg],
        \\
        V_{LS}^{s}(r,\sigma)=&-C_{\sigma}^{s}\frac{g_{s1}g_{s2}}{4\pi}\frac{u_{\sigma}^{3}}{2m_{A}m_{B}}H_{2}\Delta_{LS}.
      \end{split}
    \end{equation}
  \item Pseudoscalar mesons exchange
    \begin{equation}\label{V_P1}
      \begin{split}
        V_{SS}^{p}(r,\alpha)&=C_{\alpha}^{p}\frac{g_{p1}g_{p2}}{4\pi}\frac{u_{\alpha}^{3}}{12m_{A}m_{B}}H_{1}\Delta_{S_AS_B},
        \\
        V_{T}^{p}(r,\alpha)&=C_{\alpha}^{p}\frac{g_{p1}g_{p2}}{4\pi}\frac{u_{\alpha}^{3}}{12m_{A}m_{B}}H_{3}\Delta_{ten}.
      \end{split}
    \end{equation}
    if $u_{ex}^2=m_{ex}^2-(m_f-m_i)^2<0$, the potentials change into
    \begin{equation}\label{V_P2}
      \begin{split}
        V_{SS}^{p}(r,\alpha)&=C_{\alpha}^{p}\frac{g_{p1}g_{p2}}{4\pi}\frac{\theta_{\alpha}^{3}}{12m_{A}m_{B}}M_{1}\Delta_{S_AS_B},
        \\
        V_{T}^{p}(r,\alpha)&=C_{\alpha}^{p}\frac{g_{p1}g_{p2}}{4\pi}\frac{\theta_{\alpha}^{3}}{12m_{A}m_{B}}M_{3}\Delta_{ten},
      \end{split}
    \end{equation}
    where $\theta_{ex}^2=-[m_{ex}^2-(m_f-m_i)^2]$.
  \item Vector mesons exchange
    \begin{equation}\label{V_V}
      \begin{split}
        V_{C}^{v}(r,\beta)=&C_{\beta}^{v}\frac{u_{\beta}}{4\pi}\bigg[g_{v1}g_{v2}H_{0}+\frac{u_{\beta}^{2}}{8m_{A}m_{B}}
        \\
        &\times(g_{v1}g_{v2}+2g_{v1}f_{v2}+2g_{v2}f_{v1})H_{1}\bigg],
        \\
        V_{SS}^{v}(r,\beta)=&C_{\beta}^{v}\frac{1}{4\pi}\bigg[g_{v1}g_{v2}+g_{v1}f_{v2}+g_{V2}f_{v1}+f_{v1}f_{v2}\bigg]
        \\
        &\times\frac{u_{\beta}^{3}}{6m_{A}m_{B}}H_{1}\Delta_{S_AS_B},
        \\
        V_{T}^{v}(r,\beta)=&-C_{\beta}^{v}\frac{1}{4\pi}\bigg[g_{v1}g_{v2}+g_{v1}f_{v2}+g_{v2}f_{v1}+f_{v1}f_{v2}\bigg]
        \\
        &\times\frac{u_{\beta}^{3}}{12m_{A}m_{B}}H_{3}\Delta_{ten},
        \\
        V_{LS}^{v}(r,\beta)=&-C_{\beta}^{v}\frac{1}{4\pi}\bigg[3g_{v1}g_{v2}\Delta_{LS}+4g_{v1}f_{v2}\Delta_{LS_{A}}
        \\
        &+4g_{v2}f_{v1}\Delta_{LS_{B}}\bigg]\frac{u_{\beta}^{3}}{2m_{A}m_{B}}H_{2}.
      \end{split}
    \end{equation}
\end{itemize}
In the above expressions, the superscripts $s$, $p$ and $v$ mean
scalar, pseudoscalar and vector mesons, respectively. $\alpha=\pi,
\eta$ and $\beta=\omega, \rho, \phi$. $m_A$ and $m_B$ are the heavy
baryon masses. $g_s, g_p$ and $g_v$ are the coupling constants in
Eqs.~(\ref{cc_S}-\ref{cc_V}). 1 and 2 in the subscript are used to
mark different vertices. $C_{\sigma}^{s}$, $C_{\alpha}^{p}$ and
$C_{\beta}^{v}$ in the expressions are the isospin factors. Their
values are given in Table~\ref{Table_if}. The scalar function $H_i=H_i(\Lambda,m_{\sigma/\alpha/\beta},r)$, $M_i=M_i(\Lambda,m_{\alpha},r)$ come from Fourier transform. We give their
specific expressions in Appendix~\ref{app_func}. The subscripts $C$,
$LS$, $SS$ and $T$ denote four different kinds of potentials,
central term, spin-orbit term, spin-spin term and tensor term.
$\Delta_{S_AS_B}$, $\Delta_{LS}$ and $\Delta_{T}$ are the spin-spin
operator, spin-orbital operator and tensor operator, respectively.
Their specific forms are collected in Appendix~\ref{app_matrix}.

Apart from the two baryon systems, we also calculate the possible
molecular states with one baryon and one antibaryon. We use the
G-parity rule to derive the potential between a baryon and its
antibaryon. The potentials in Eqs.~(\ref{V_S}-\ref{V_V}) still hold
up to an extra factor $(-1)^{I_G}$, where $I_G$ is the G-parity of
the exchanged meson. The extra factor is absorbed into the isospin
factor of baryon-antibaryon system in Table~\ref{Table_if}.
\begin{table*}[htp]
 \centering
 \caption{The isospin factors for two baryon systems and baryon-antibaryon systems. The factors $(-1)^{I_G}$ from G-parity rule have been absorbed by the isospin factors in the right panel.}\label{Table_if}
\begin{tabular}{ccccccc|ccccccc}
\hline \hline ~~States~~ & ~~$C_{\sigma}^{s}$~~ & ~~$C_{\pi}^{p}$~~
& ~~$C_{\eta}^{p}$~~ & ~~$C_{\rho}^{v}$~~ & ~~$C_{\omega}^{v}$~~ &
~~$C_{\phi}^{v}$~~ & ~~States~~ & ~~$C_{\sigma}^{s}$~~ &
~~$C_{\pi}^{p}$~~ & ~~$C_{\eta}^{p}$~~ & ~~$C_{\rho}^{v}$~~ &
~~$C_{\omega}^{v}$~~ & ~~$C_{\phi}^{v}$~~\tabularnewline \hline
$\Sigma_{c}^{*}\Sigma_{c}^{*}[I=0]$ & 1 & -1 & 1/6 & -1 & 1/2 & 0 &
$\Sigma_{c}^{*}\bar{\Sigma}_{c}^{*}[I=0]$ & 1 & 1 & 1/6 & -1 & -1/2
& 0\tabularnewline $\Sigma_{c}^{*}\Sigma_{c}^{*}[I=1]$ & 1 & -1/2 &
1/6 & -1/2 & 1/2 & 0 & $\Sigma_{c}^{*}\bar{\Sigma}_{c}^{*}[I=1]$ & 1
& 1/2 & 1/6 & -1/2 & -1/2 & 0\tabularnewline
$\Sigma_{c}^{*}\Sigma_{c}^{*}[I=2]$ & 1 & 1/2 & 1/6 & 1/2 & 1/2 & 0
& $\Sigma_{c}^{*}\bar{\Sigma}_{c}^{*}[I=2]$ & 1 & -1/2 & 1/6 & 1/2 &
-1/2 & 0\tabularnewline $\Xi_{c}^{*}\Xi_{c}^{*}[I=0]$ & 1 & -3/8 &
1/24 & -3/8 & 1/8 & 1/4 & $\Xi_{c}^{*}\bar{\Xi}_{c}^{*}[I=0]$ & 1 &
3/8 & 1/24 & -3/8 & -1/8 & -1/4\tabularnewline
$\Xi_{c}^{*}\Xi_{c}^{*}[I=1]$ & 1 & 1/8 & 1/6 & 1/8 & 1/8 & 1/4 &
$\Xi_{c}^{*}\bar{\Xi}_{c}^{*}[I=1]$ & 1 & -1/8 & 1/6 & 1/8 & -1/8 &
-1/4\tabularnewline $\Omega_{c}^{*}\Omega_{c}^{*}[I=0]$ & 1 & 0 &
2/3 & 0 & 0 & 1 & $\Omega_{c}^{*}\bar{\Omega}_{c}^{*}[I=0]$ & 1 & 0
& 2/3 & 0 & 0 & -1\tabularnewline \hline \hline
\end{tabular}
\end{table*}

For the molecular states composed of two spin-$\frac{3}{2}$ baryons,
the total spin $J$ can be 0, 1, 2 and 3. The wave function of bound
states in S-wave reads
\begin{equation}
  \Psi(r,\theta,\phi)\chi_{ss_z}=T(r)\ket{{}^{2S+1}S_J}.
\end{equation}
For the $J=0$ and $1$ systems, we also take the couple channel
effect from systems with higher orbital angular momentum into
consideration.  For the $J=0$ states, we consider the S-D wave
mixing. The wave function reads
\begin{equation}
  \Psi(r,\theta,\phi)^T\chi_{ss_z}^T=\left[\begin{array}{c}
T_{S}(r)\\
0
\end{array}\right]\ket{{}^{1}S_{0}}+\left[\begin{array}{c}
0\\
T_{D}(r)
\end{array}\right]\ket{{}^{5}D_{0}},
\end{equation}
where $T_i$ means the radial wave functions for different channels.
For the $J=1$ states, we consider the G-wave mixing additionally.
The wave function reads

%\begin{widetext}
\begin{equation}
\begin{split}
\Psi(r,\theta,\phi)^T\chi_{ss_z}^T=&\left[\begin{array}{c}
T_{S}(r)\\
0\\
0\\
0
\end{array}\right]\ket{{}^{3}S_{1}}+\left[\begin{array}{c}
0\\
T_{D}(r)\\
0\\
0
\end{array}\right]\ket{{}^{3}D_{1}}
\\
&+\left[\begin{array}{c}
0\\
0\\
T_{D}'(r)\\
0
\end{array}\right]\ket{{}^{7}D_{1}}+\left[\begin{array}{c}
0\\
0\\
0\\
T_{G}(r)
\end{array}\right]\ket{{}^{7}G_{1}}.
\end{split}
\end{equation}
%\end{widetext}

The matrix elements of operators in Eqs.~(\ref{V_S}-\ref{V_V}) can
be derived explicitly,
\begin{itemize}
\item Single channel
\begin{equation}
\begin{split}
&\Delta_{LS}=0,~\Delta_{LS_A}=0,~\Delta_{LS_B}=0,~\Delta_{T}=0,
\\
&\Delta_{S_AS_B}=\left(2S(S+1)-15\right)/9.
\end{split}
\end{equation}
\item Couple channel for $J^P=0^+$
\begin{equation}
\begin{split}
\Delta_{LS}&=\left[\begin{array}{cc}
0 & 0\\
0 & -2
\end{array}\right],~
\Delta_{LS_A}=\left[\begin{array}{cc}
0 & 0\\
0 & -1
\end{array}\right],
\\
\Delta_{LS_B}&=\left[\begin{array}{cc}
0 & 0\\
0 & -1
\end{array}\right],~
\Delta_{S_AS_B}=\left[\begin{array}{cc}
-\frac{5}{3} & 0\\
0 & -1
\end{array}\right],
\\
\Delta_{T}&=\left[\begin{array}{cc}
0 & -\frac{4}{3}\\
-\frac{4}{3} & -\frac{4}{3}
\end{array}\right].
\end{split}
\end{equation}
\item Couple channel for $J^P=1^+$
\begin{equation}
\begin{split}
\Delta_{LS}&=\left[\begin{array}{cccc}
0 & 0 & 0 & 0\\
0 & -1 & 0 & 0\\
0 & 0 & -\frac{8}{3} & 0\\
0 & 0 & 0 & -5
\end{array}\right],~
\Delta_{LS_A}=\left[\begin{array}{cccc}
0 & 0 & 0 & 0\\
0 & -\frac{1}{2} & 0 & 0\\
0 & 0 & -\frac{4}{3} & 0\\
0 & 0 & 0 & -\frac{5}{2}
\end{array}\right],
\\
\Delta_{LS_B}&=\left[\begin{array}{cccc}
0 & 0 & 0 & 0\\
0 & -\frac{1}{2} & 0 & 0\\
0 & 0 & -\frac{4}{3} & 0\\
0 & 0 & 0 & -\frac{5}{2}
\end{array}\right],
\Delta_{S_AS_B}=\left[\begin{array}{cccc}
-\frac{11}{9} & 0 & 0 & 0\\
0 & -\frac{11}{9} & 0 & 0\\
0 & 0 & 1 & 0\\
0 & 0 & 0 & 1
\end{array}\right],
\\
\Delta_{T}&=\left[\begin{array}{cccc}
0 & \frac{34\sqrt{2}}{45} & -\frac{4\sqrt{7}}{15} & 0\\
\frac{34\sqrt{2}}{45} & -\frac{34}{45} & \frac{4\sqrt{14}}{105} & -\frac{4\sqrt{42}}{35}\\
-\frac{4\sqrt{7}}{15} & \frac{4\sqrt{14}}{105} & -\frac{48}{35} & \frac{4\sqrt{3}}{35}\\
0 & -\frac{4\sqrt{42}}{35} & \frac{4\sqrt{3}}{35} & -\frac{10}{7}
\end{array}\right].
\end{split}
\end{equation}
\end{itemize}

The derivation details about these matrix elements of the operators
can also be found in Appendix~\ref{app_matrix}.

%%%%%%%%%%%%%%%%%%%%%%%
\section{NUMERICAL RESULTS }\label{sec_num}
%%%%%%%%%%%%%%%%%%%%%%%

With the effective potential, we solve the Schr\"odinger equation
numerically and then obtain the binding energy and radial wave
function. We can calculate the root-mean-square radius ($R_{rms}$)
with the radial wave function, which can help us to check the
self-consistency and rationality of the molecular state. The
root-mean-square radius is
\begin{equation}
  R_{rms}^2=\int \sum_i T_i(r)T^*_i(r) r^4dr,
\end{equation}
where $T_i$ is the radial wave function of channel $i$. The $\sum$
means the sum of all different channels. We can also calculate the
individual probability for each channel.
\begin{equation}
  P_{T_i}=\int T_i^*(r)T_i(r)r^2dr.
\end{equation}

In our results we keep one decimal of energies and root-mean square radii, which dose not represent our accuracy. The numbers are simply numerical results under this framework. The actual uncertainty stemming from the theoretical framework may be quite large.

%%%%%%%%%%%%%%%%%%%%%
%%%%%%%%%%%%%%%%%%%%%%%%

We calculate the possible molecular states
formed by two baryons.
The total wave function of the two baryons
system is antisymmetric for the Pauli Principle.
Since the spatial wave function is symmetric, the $S=0,2$ state has the symmetric isospin wave function and $S=1,3$ state has the antisymmetric isospin wave function. We also calculate the possible molecular states composed of one baryon and one antibaryon.
Since a baryon-antibaryon molecular state may decay into three mesons through quark rearrangement, which renders the bound states unstable.
The binding solution in our calculations for the baryon-antibaryon system may be a candidate of the molecule-type resonance.
Thus, the numerical results of the baryon-antibaryon systems are collected in Appendix~\ref{app_BBbarsc}.

\subsection{Single Channel Calculation}\label{subsubsec_BB_sc}

\begin{table*}[!htb]
 \centering
 \caption{The numerical results for two charmed baryons in single channel calculation. $\Lambda $ is the cutoff parameter. "$E$" is the binding energy. $R_{rms}$ is the root-mean-square radius. We use $[I(J^P)]$ to mark different states. The sign ``$\times $" means no reasonable binding solution.}\label{Table_num_BB_sc}
\begin{tabular}{cccc|cccc}
\hline \hline
States & $\Lambda$(MeV) & E(MeV) & $R_{rms}$(fm) & States & $\Lambda$(MeV) & E(MeV) & $R_{rms}$(fm)\tabularnewline
\hline
\multirow{3}{*}{$\Sigma_{c}^{*}\Sigma_{c}^{*}[0(0^{+})]$} &  &  &  & \multirow{3}{*}{$\Xi_{c}^{*}\Xi_{c}^{*}[0(1+)]$} & 850 & 10.3 & 1.2\tabularnewline
 &  & $\times$ &  &  & 900 & 40.0 & 0.7\tabularnewline
 &  &  &  &  & 950 & 85.9 & 0.5\tabularnewline
\hline
\multirow{3}{*}{$\Sigma_{c}^{*}\Sigma_{c}^{*}[0(2^{+})]$} & 800 & 2.6 & 2.0 & \multirow{3}{*}{$\Xi_{c}^{*}\Xi_{c}^{*}[0(3+)]$} & 1100 & 2.1 & 2.5\tabularnewline
 & 850 & 11.5 & 1.1 &  & 1200 & 6.1 & 1.7\tabularnewline
 & 900 & 28.1 & 0.8 &  & 1300 & 11.0 & 1.4\tabularnewline
\hline
\multirow{3}{*}{$\Sigma_{c}^{*}\Sigma_{c}^{*}[2(0^{+})]$} & 1400 & 2.3 & 2.7 & \multirow{3}{*}{$\Xi_{c}^{*}\Xi_{c}^{*}[1(0+)]$} & 1300 & 1.2 & 3.2\tabularnewline
 & 1500 & 5.8 & 1.9 &  & 1400 & 2.9 & 2.2\tabularnewline
 & 1600 & 11.1 & 1.6 &  & 1500 & 5.6 & 1.8\tabularnewline
\hline
\multirow{3}{*}{$\Sigma_{c}^{*}\Sigma_{c}^{*}[2(2^{+})]$} &  &  &  & \multirow{3}{*}{$\Xi_{c}^{*}\Xi_{c}^{*}[1(2+)]$} & 1100 & 2.7 & 2.1\tabularnewline
 &  & $\times$ &  &  & 1300 & 4.2 & 1.8\tabularnewline
 &  &  &  &  & 1500 & 3.8 & 1.8\tabularnewline
\hline
\multirow{3}{*}{$\Sigma_{c}^{*}\Sigma_{c}^{*}[1(1^{+})]$} & 800 & 18.8 & 0.9 & \multirow{3}{*}{$\Omega_{c}^{*}\Omega_{c}^{*}[0(0^{+})]$} & 1100 & 2.4 & 2.3\tabularnewline
 & 850 & 35.0 & 0.8 &  & 1200 & 3.5 & 2.0\tabularnewline
 & 900 & 55.6 & 0.6 &  & 1300 & 6.0 & 1.7\tabularnewline
\hline
\multirow{3}{*}{$\Sigma_{c}^{*}\Sigma_{c}^{*}[1(3^{+})]$} & 1300 & 3.0 & 2.2 & \multirow{3}{*}{$\Omega_{c}^{*}\Omega_{c}^{*}[0(2^{+})]$} & 1000 & 5.2 & 1.6\tabularnewline
 & 1400 & 5.9 & 1.7 &  & 1200 & 11.8 & 1.2\tabularnewline
 & 1500 & 9.5 & 1.4 &  & 1400 & 5.9 & 1.6\tabularnewline
\hline \hline
\end{tabular}
\end{table*}

We first perform the single channel calculation to find the possible
molecular states. Here we calculate the S-wave systems. We give the binding energies and the root-mean-square radii of possible
molecular states in Table~\ref{Table_num_BB_sc}.

\begin{figure*}[htp]
\centering
\includegraphics[width=0.95\textwidth]{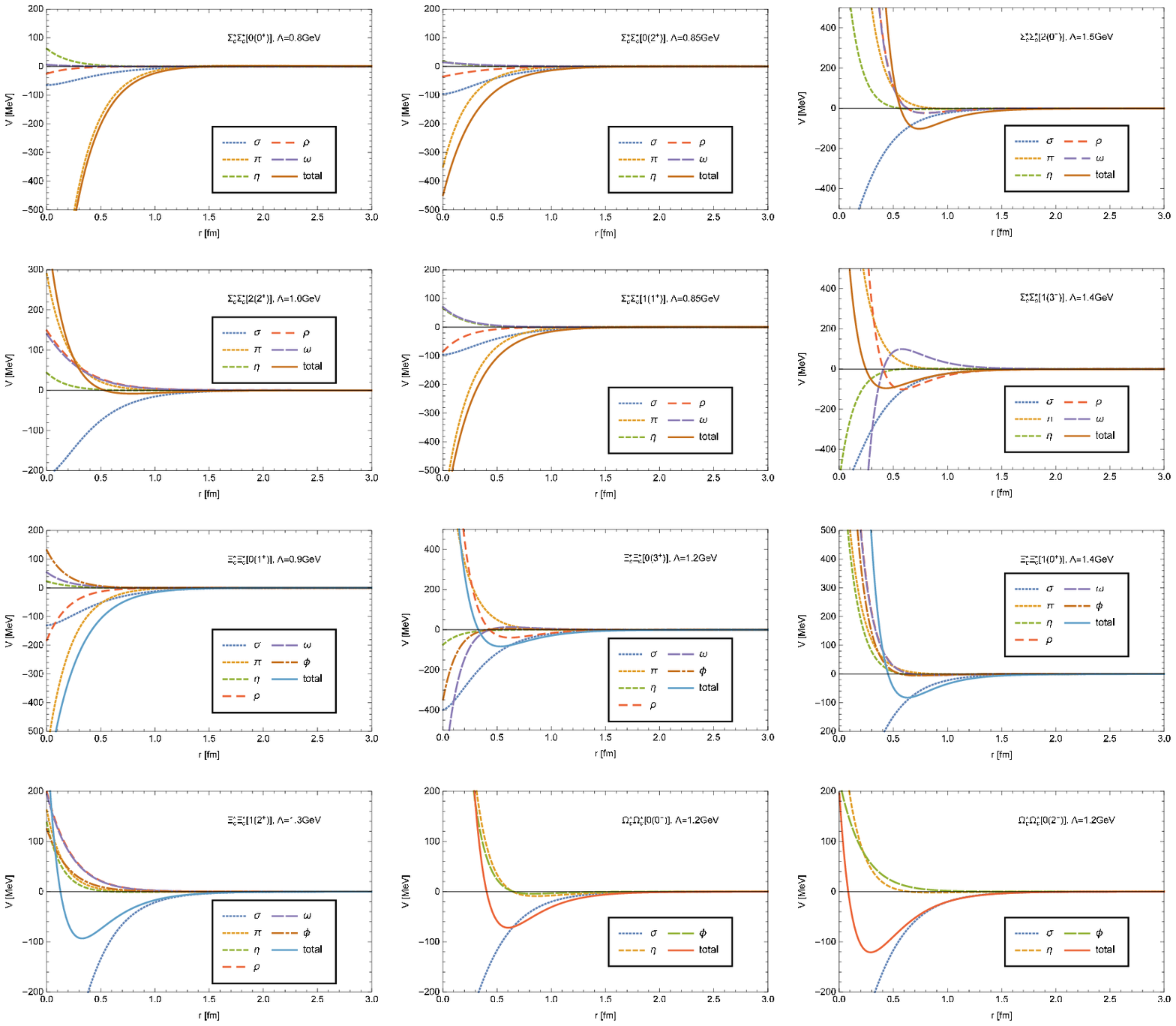}
\caption{The interaction potentials for the two charmed baryons in
the S-wave.}\label{fig_p_BB_sc}
\end{figure*}

Their potentials
are shown in Fig.~\ref{fig_p_BB_sc}. There exist binding solutions
for the $\Sigma_{c}^{*}\Sigma_{c}^{*}[0(2^{+})$, $2(0^{+})$, $1(1^{+})$, $1(3^{+})]$,
$\Xi_{c}^{*}\Xi_{c}^{*}[0(1^{+})$, $0(3^{+})$, $1(0^{+})$, $1(2^{+})]$, and
$\Omega_{c}^{*}\Omega_{c}^{*}[0(0^{+})$, $0(2^{+})]$ systems. All
these bound states are good molecule candidates. Each of them has a
small binding energy and suitable root-mean-square radius under a
reasonable range of the cutoff parameter.

There are four candidates of $\Sigma_{c}^{*}\Sigma_{c}^*$ molecular states.
For the $\Sigma_{c}^{*}\Sigma_{c}^{*}[0(2^{+})]$ system, the $\pi$ exchange potential is attractive when $r < 1$ fm, which provides the main part of the total potential. The binding energy is 2.6-28.1 MeV when the cutoff parameter varies from 0.8 GeV to 0.9 GeV. For the $\Sigma_{c}^{*}\Sigma_{c}^{*}[2(0^{+})]$ system, $\pi$, $\eta$, $\rho$ and $\omega$ exchange potentials are considerably repulsive in the short-range and become attractive when $r> 0.6$ fm. The $\sigma$ exchange potential is always attractive. As a result, the total potential is slightly attractive in the range $0.7 < r < 1.5$ fm. A bound state appears with binding energy about 2.3-11.1 MeV when the cutoff parameter is around 1.4-1.6 GeV.
The potential of $\Sigma_{c}^{*}\Sigma_{c}^{*}[1(1^{+})]$ system is similar to that of $[0(2^+)]$ system. The binding energy of the state is 18.8-55.6 MeV, while the cutoff parameter is 0.8-0.9 GeV. For the $\Sigma_{c}^{*}\Sigma_{c}^{*}[1(3^{+})]$ system, the total potential is repulsive when $r < 0.3$ fm. In the range $0.4 < r < 1.5$ fm, the contributions of the $\rho$ and $\omega$ exchange cancel with each other
significantly, which makes the total potential weakly attractive. As
a result there exists a weak binding solution with the cutoff parameter is around 1.3-1.5 GeV.

The potentials of the $\Xi_{c}^{*}\Xi_{c}^{*}$ systems can be slightly attractive with an appropriate cutoff. For the $\Xi_{c}^{*}\Xi_{c}^{*}[0(1^{+})]$ system, the attractive potential arising from the one pion exchange leads to a binding solution. The binding energy is 10.3-85.9 MeV while the cutoff parameter is 0.85-0.95 GeV. For the $\Xi_{c}^{*}\Xi_{c}^{*}[0(3^+)]$ systems, the $\pi$ and $\sigma$ exchange potentials are attractive around 0.5-1.0 fm. As a result, a slightly bound state with binding energy 2.1-11.0 MeV appears when the cutoff parameter is 1.1-1.3 GeV. For the $\Xi_{c}^{*}\Xi_{c}^{*}[1(0^+),1(2^+)]$, the attractive parts of the potentials mainly come from the $\sigma$ exchange. The binding energy of the $\Xi_{c}^{*}\Xi_{c}^{*}[1(0^+)]$ around 1.2-5.6 MeV with the cutoff is 1.3-1.5 GeV. The binding energy of the $\Xi_{c}^{*}\Xi_{c}^{*}[1(2^+)]$ is 2.7-3.8 MeV when the cutoff varies from 1.1 GeV to 1.5 GeV.

For the $\Omega_{c}^{*}\Omega_{c}^{*}$ systems, two loosely bound
states are obtained. There dose not exist the $\pi$ exchange between two
$\Omega_{c}^{*}$s, which usually provides the main part of the total
potential. Even so, the $\sigma$, $\eta$ and $\phi$ exchanges can
also lead to a weakly attractive potential around 1 fm. For the
$\Omega_{c}^{*}\Omega_{c}^{*}[0(0^{+})]$ system, we find a bound
state with the binding energy about 2.4-6.0 MeV when the cutoff
is 1.1-1.3 GeV. And for the $\Omega_{c}^{*}\Omega_{c}^{*}[0(2^{+})]$
system, a bound state with binding energy 5.2-11.8 MeV appears
when the cutoff parameter is 1.0-1.4 GeV.

From the Fig.~\ref{fig_p_BB_sc}, one can notice that a strong attraction exists for the systems $\Sigma_{c}^{*}\Sigma_{c}^{*}[0(0^{+})]$ in the range $r < 1$ fm. The strong attractive potential is provided by the $\pi$ exchange. The contribution of the other meson exchange is quite small. The strong attractive total potential generates a tightly bound system. We get a binding solution with very large binding energy and very small root-mean-square radius. The strong attraction in the channel strongly indicates that there may exist the heavy analogue of the H-dibaryon with the configurations such as ccqqqq where q denotes the up or down quark. For the system $\Sigma_{c}^{*}\Sigma_{c}^{*}[2(2^{+})]$, the total attractive potential is too weak to form a bound state. Actually we find no binding solution in a reasonable range for the cutoff parameter.

\subsection{Couple Channel Calculation}\label{subsubsec_BB_cc}

\begin{figure*}[htp]
\centering
\includegraphics[width=0.95\textwidth]{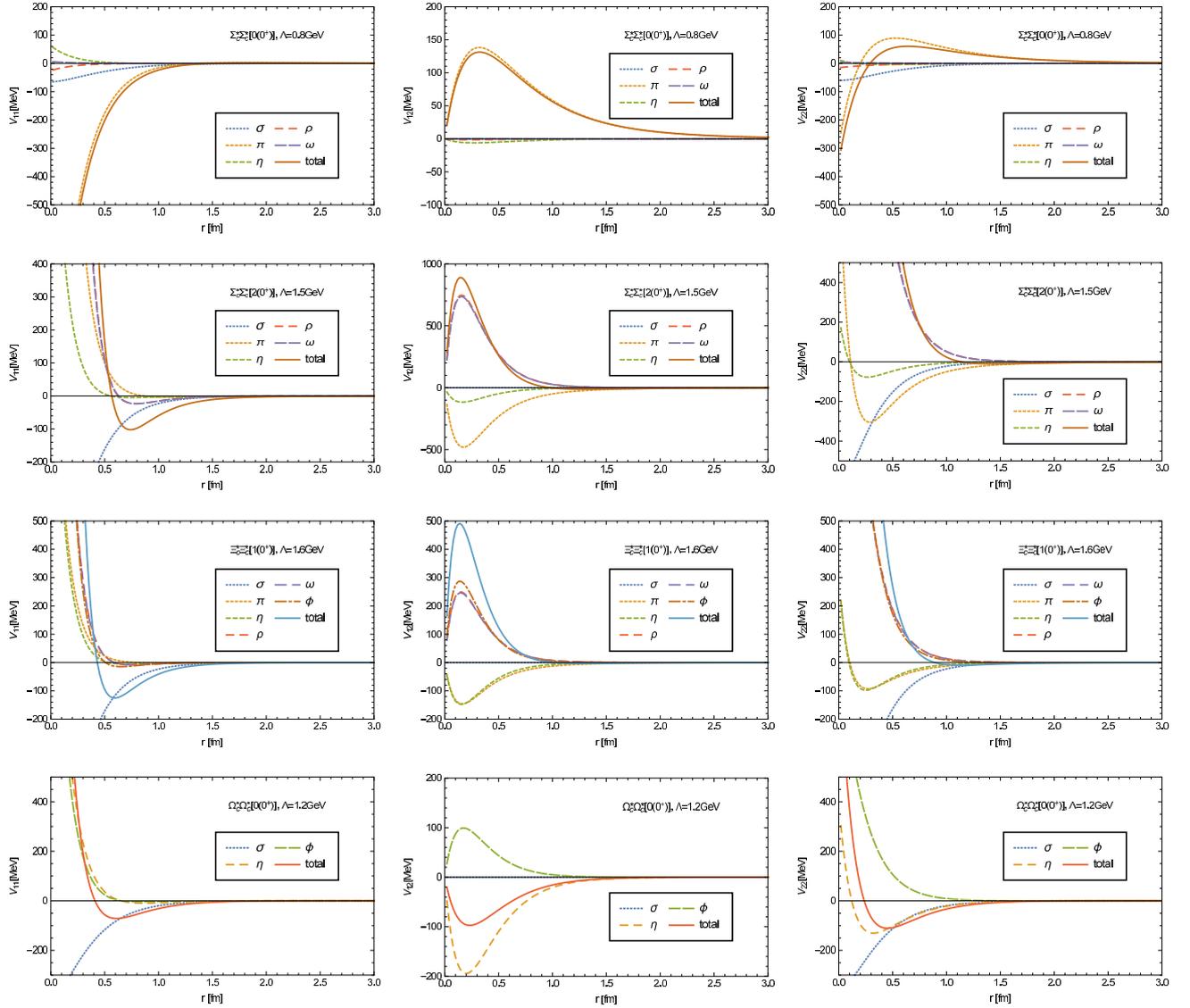}
\caption{The interaction potentials for the two charmed baryons
with total spin 0. Two channels are included. $V_{11},V_{12}$ and
$V_{22}$ denote the ${}^{1}S_{0}\leftrightarrow{}^{1}S_{0}$,
${}^{1}S_{0}\leftrightarrow{}^{5}D_{0}$ and
${}^{5}D_{0}\leftrightarrow{}^{5}D_{0}$ transitions potentials. The
four rows from top to bottom are for $
\Sigma_{c}^{*}\Sigma_{c}^{*}[0(0^{+})],
\Sigma_{c}^{*}\Sigma_{c}^{*}[2(0^{+})],\Xi_{c}^{*}\Xi_{c}^{*}[1(0^{+})]$
and $\Omega_{c}^{*}\Omega_{c}^{*}[0(0^{+})]$.}\label{fig_p_BB_cc1}
\end{figure*}

\begin{figure*}[htp]
\centering
\includegraphics[width=0.95\textwidth]{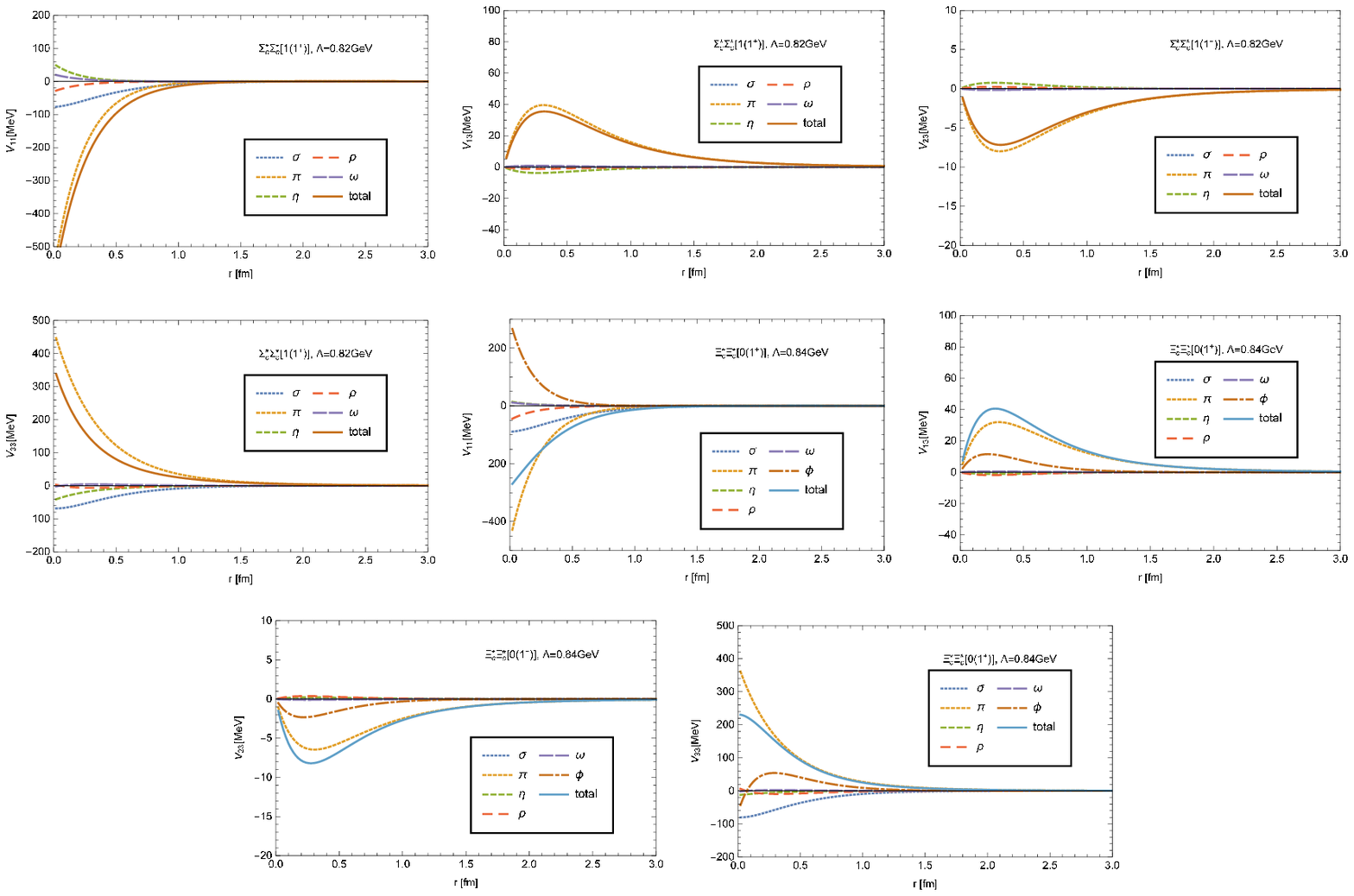}
\caption{The interaction potentials for the
$\Sigma_c^{*}\Sigma_c^{*}$ and $\Xi_c^{*}\Xi_c^{*}$ systems with
total spin 1 in couple channel calculation. The subscript number
``1-4" of ``$V$" means states${}^3S_{1},{}^3D_{1},{}^7D_{1}$ and
${}^7G_{1}$ in sequence. We show only four representative potentials
here. $V_{22}$ are similar to $V_{11}$. $V_{24}$ is
similar to $V_{13}$. $V_{44}$ are similar to
$V_{33}$.}\label{fig_p_BB_cc2}
\end{figure*}

Here, we consider the couple channel effect between states with
different spin and angular momentum for comparison. These states are
mixed by the tensor operator. For the system with spin 0, we
consider the S-D wave mixing. For the system with spin 1, we add
G-wave besides the S- and D-waves. For the D-wave channel, the spin of two
baryons can be 1 or 3. The numerical results including the binding
energy, root-mean-square radius and the percentages of different
channels are shown in Table~\ref{Table_num_BB_cc1} and Table~\ref{Table_num_BB_cc2}. The potentials of different channels are given in
Figs.~\ref{fig_p_BB_cc1} and \ref{fig_p_BB_cc2} respectively. There are three
good candidates of molecular systems with total spin 0,
$\Sigma_{c}^{*}\Sigma_{c}^{*}[2(0^{+})]$,
$\Xi_{c}^{*}\Xi_{c}^{*}[1(0^{+})]$, and
$\Omega_{c}^{*}\Omega_{c}^{*}[0(0^{+})]$.
For the systems with total spin 1, the $\Sigma_{c}^{*}\Sigma_{c}^{*}[1(1^{+})]$ and $\Xi_{c}^{*}\Xi_{c}^{*}[0(1^{+})]$ are also good candidates of molecular states.

\begin{table*}[htp]
\centering \caption{The numerical results for two charmed baryons
with total spin 0 in couple channel calculation. $\Lambda $ is the
cutoff parameter. "$E$" is the binding energy. $R_{rms}$ is the
root-mean-square radius. We use $[I(J^P)]$ to mark different states.
$P_{S}$ is the percentage of the S wave, and $P_{D}$ is the
percentage of the D wave.}\label{Table_num_BB_cc1}
\begin{tabular}{cccccc|cccccc}
\hline \hline
States & $\Lambda$(MeV) & E(MeV) & $R_{rms}$(fm) & $P_{S}(\%)$ & $P_{D}(\%)$ & States & $\Lambda$(MeV) & E(MeV) & $R_{rms}$(fm) & $P_{S}(\%)$ & $P_{D}(\%)$\tabularnewline
\hline
\multirow{3}{*}{$\Sigma_{c}^{*}\Sigma_{c}^{*}[0(0^{+})]$} &  &  &  &  &  & \multirow{3}{*}{$\Xi_{c}^{*}\Xi_{c}^{*}[1(0^{+})]$} & 1400 & 3.1 & 2.2 & 99.9 & 0.1\tabularnewline
 &  &  & $\times$ &  &  &  & 1600 & 11.5 & 1.4 & 99.6 & 0.4\tabularnewline
 &  &  &  &  &  &  & 1800 & 27.8 & 1.1 & 98.8 & 1.2\tabularnewline
\hline
\multirow{3}{*}{$\Sigma_{c}^{*}\Sigma_{c}^{*}[2(0^{+})]$} & 1400 & 3.2 & 2.7 & 98.5 & 1.5 & \multirow{3}{*}{$\Omega_{c}^{*}\Omega_{c}^{*}[0(0^{+})]$} & 1000 & 2.1 & 2.5 & 98.5 & 1.5\tabularnewline
 & 1500 & 7.5 & 2.0 & 99.0 & 0.1 &  & 1200 & 6.0 & 1.8 & 97.7 & 2.3\tabularnewline
 & 1600 & 14.8 & 1.6 & 99.3 & 0.7 &  & 1400 & 11.3 & 1.5 & 99.5 & 0.5\tabularnewline
\hline \hline
\end{tabular}
\end{table*}

\begin{table*}[htp]
\centering \caption{The numerical results for two charmed baryons
with total spin 1 in couple channel calculation. $\Lambda $ is the
cutoff parameter. "$E$" is the binding energy. $R_{rms}$ is the
root-mean-square radius. We use $[I(J^P)]$ to mark different states.
$P_{S}$ is the percentage of the ${}^3S_1$, $P_{D1}$ is the
percentage of the ${}^3D_1$, $P_{D2}$ is the percentage of the
${}^7D_1$, and $P_{G}$ is the percentage of the
${}^7G_1$.}\label{Table_num_BB_cc2}
\begin{tabular}{cccccccc}
\hline \hline
States & $\Lambda$(MeV) & E(MeV) & $R_{rms}$(fm) & $P_{S}(\%)$ & $P_{D1}(\%)$ & $P_{D2}(\%)$ & $P_{G}(\%)$\tabularnewline
\hline
\multirow{3}{*}{$\Sigma_{c}^{*}\Sigma_{c}^{*}[1(1^{+})]$} & 800 & 23.3 & 1.1 & 98.1 & 1.5 & 0.4 & 0.0\tabularnewline
 & 820 & 29.7 & 1.0 & 98.1 & 1.5 & 0.4 & 0.0\tabularnewline
 & 840 & 36.8 & 0.9 & 98.2 & 1.4 & 0.4 & 0.0\tabularnewline
\hline
\multirow{3}{*}{$\Xi_{c}^{*'}\Xi_{c}^{*'}[0(1+)]$} & 820 & 4.9 & 1.9 & 97.4 & 2.0 & 0.6 & 0.0\tabularnewline
 & 840 & 11.0 & 1.5 & 97.5 & 1.9 & 0.6 & 0.0\tabularnewline
 & 860 & 19.6 & 1.2 & 97.9 & 1.6 & 0.5 & 0.0\tabularnewline
\hline \hline
\end{tabular}
\end{table*}

For the $\Sigma_{c}^{*}\Sigma_{c}^{*}[2(0^{+})]$ system, one notices that the interaction potentials of the ${}^5D_0$ channel and the transition potential of ${}^{1}S_{0}\leftrightarrow{}^{5}D_{0}$ in Fig.~\ref{fig_p_BB_cc1} are both repulsive. There exists a loosely bound state with binding energy 3.2-14.8 MeV, while the cutoff is 1.4-1.6 GeV. The S-D mixing is quite small for the system. From Table~\ref{Table_num_BB_cc1}, we notice that the probability of the D-wave is about 2\%. For the $\Xi_{c}^{*}\Xi_{c}^{*}[1(0^{+})]$ and the $\Omega_{c}^{*}\Omega_{c}^{*}[0(0^{+})]$ systems, the S-D transition potentials both affect the solutions slightly. Their binding energies change slightly compared with the single channel cases. For the $\Sigma_{c}^{*}\Sigma_{c}^{*}[0(0^{+})]$ system, there is no slightly bound solution even if we consider the S-D wave mixing.

For the $\Sigma_{c}^{*}\Sigma_{c}^{*}[1(1^{+})]$ and $\Xi_{c}^{*}\Xi_{c}^{*}[0(1^{+})]$ systems with $J=1$, we can still find loosely bound solutions when the channel mixing effect is considered. Compared with the single channel cases, their binding energies become slightly larger, and more dependent on the cutoff parameter. We show their potentials in the Fig.~\ref{fig_p_BB_cc2}.

%%%%%%%%%%%%%%%%%%%%%
%%%%%%%%%%%%%%%%%%%%%%%%

%%%%%%%%%%%%%%%%%%%%%%%%%%%%%%%%%%
\section{DISCUSSIONS AND CONCLUSIONS}\label{sec_dis}
%%%%%%%%%%%%%%%%%%%%%%%%%%%%%%%%%%

In this work, we have performed a systematic investigation of the
possible deuteron-like molecules composed of a pair of
spin-$\frac{3}{2}$ singly charmed baryons. We have calculated the single channel
results for all possible states with different total spins, and
considered the couple channel effect for total spin 0 and 1 systems.
For the systems with total spin 0, the channel mixing is between
${}^1S_0$ and ${}^5D_0$. For the systems with total spin 1, we
include four channels, ${}^3S_1$, ${}^3D_1$, ${}^7D_1$ and ${}^7G_1$
in calculation.

The hadronic molecule is assumed to be a loosely bound state of two color singlet components. The formation of the molecular state mainly arises from the relatively long range attraction, which can be described well in OBE model. However, the extremely short range interaction from the OBE model may be not very convincing. Therefore, the very deep binding solution arising from the strong short attraction may be not physical. Thus, according to an empirical and intuitive approach suggested in
Ref.~\cite{Li:2012bt}, the binding energy of a molecular state
formed by two charmed baryons is expected to be less than 240 MeV,
and the root-mean-square radius to be larger than 0.6-1.0 fm. With
the help of the criteria, we have used the binding energy and
root-mean-square radius to make some educated guesses whether the
system is a loosely bound state. We use $\times$ in the table of results to denote those systems with very large binding energies.

For the ten systems, $\Sigma_{c}^{*}\Sigma_{c}^{*}$$[0(2^{+})$, $2(0^{+})$,
$1(1^{+})$, $1(3^{+})]$, $\Xi_{c}^{*}\Xi_{c}^{*}$$[0(1^+)$, $0(3^+)$, $1(0^+)$,
$1(2^+)]$ and $\Omega_{c}^{*}\Omega_{c}^{*}$$[0(0^+)$, $0(2^+)]$, we have
obtained loosely bound solutions with small binding energies and
appropriate sizes. After we consider the channel mixing effect, the loosely binding solutions of the $\Sigma_{c}^{*}\Sigma_{c}^{*}[2(0^{+})]$, $[1(1^{+})]$, $\Xi_{c}^{*}\Xi_{c}^{*}[1(0^+)]$, $[0(1^+)]$, $\Omega_{c}^{*}\Omega_{c}^{*}$$[0(0^+)]$ systems still exist. The multichannel effect always makes the binding slightly deeper. The cutoff parameters of the systems are almost in the
range of 0.8-1.5 GeV, while that for the $\Sigma_{c}^{*}\Sigma_{c}^{*}$$[2(0^{+})]$ system is a little larger. Although the experience of the deuteron suggests a range from 0.8 GeV to 1.5 GeV, it may be reasonable to slightly widen the range for
a much heavier system. Thus, they are all good molecular candidates. For the $\Sigma_{c}^{*}\Sigma_{c}^{*}[2(2^{+})]$ system, the potential is hardly attractive, and we find no binding solution with a reasonable cutoff parameter. For the $\Sigma_{c}^{*}\Sigma_{c}^{*}[0(0^{+})$ system, the OBE potential is strongly attractive, which indicates that the tightly bound heavy dibaryon may exist with the configurations such as ccssqq or ccqqqq where q denotes the up or down quark.

The mass difference between $\Omega_c^*$ and $\Omega_c$ is too small
for any strong decay to occur. As a result, the $\Omega_c^*$ decays mainly
via $\Omega_c^*\rightarrow\Omega_c\gamma$. Once the states are produced in experiments, they would be stable. Thus the $\Omega_c^*\Omega_c^*$ systems may be observed in the future.

We also calculate the systems formed by one baryon and one antibaryon in Appendix~\ref{app_BBbarsc}.
The present formalism can be extended easily to the loosely bound
systems composed of two different spin-$\frac{3}{2}$ baryons, only
by adding the influence of the $K$ and $K^*$ exchanges. The
framework can be used to study the systems composed of one spin
$\frac{1}{2}$-baryon and one spin-$\frac{3}{2}$ baryon. One may also
extend the couple channel effect to the systems with different
particles, such as $\Sigma_c^*\Sigma_c^*[0(0^+)]\leftrightarrow
\Sigma_c\Sigma_c[0(0^+)]$.

%%%%%%%%%%%%%%%%%%%
\bigskip\noindent\textbf{Acknowledgements:}
BY is very grateful to X.Z Weng, H.S. Li and G.J. Wang for very
helpful discussions. This project is supported by the National
Natural Science Foundation of China under Grants 11575008,
11621131001 and National Key Basic Research Program of China
(2015CB856700).

\begin{appendix}

%%%%%%%%%%%%%%%%%%%%%%%%%%%
\section{Expressions of Special Functions and Some Fourier Transformation Formulae} \label{app_func}
%%%%%%%%%%%%%%%%%%%%%%%%%%%%

The definitions of $H_i$ etc. are

\begin{equation}
\begin{split}
H_{0}(\Lambda,m,r)&=Y(ur)-\frac{\lambda}{u}Y(\lambda
r)-\frac{r\beta^{2}}{2u}Y(\lambda r),
\\
H_{1}(\Lambda,m,r)&=Y(ur)-\frac{\lambda}{u}Y(\lambda
r)-\frac{r\lambda^{2}\beta^{2}}{2u^{3}}Y(\lambda r),
\\
H_{2}(\Lambda,m,r)&=Z_{1}(ur)-\frac{\lambda^{3}}{u^{3}}Z_{1}(\lambda
r)-\frac{\lambda\beta^{2}}{2u^{3}}Y(\lambda r),
\\
H_{3}(\Lambda,m,r)&=Z(ur)-\frac{\lambda^{3}}{u^{3}}Z(\lambda
r)-\frac{\lambda\beta^{2}}{2u^{3}}Z_{2}(\lambda r),
\\
M_{0}(\Lambda,m,r)&=-\frac{1}{\theta r}\left[\cos(\theta
r)-e^{-\lambda r}\right]+\frac{\beta^{2}}{2\theta\lambda}e^{-\lambda
r},
\\
M_{1}(\Lambda,m,r)&=-\frac{1}{\theta r}\left[\cos(\theta
r)-e^{-\lambda
r}\right]-\frac{\lambda\beta^{2}}{2\theta^{3}}e^{-\lambda r},
\\
M_{3}(\Lambda,m,r)&=-\left[\cos(\theta r)-\frac{3\sin(\theta
r)}{\theta r}-3\frac{\cos(\theta
r)}{\theta^{2}r^{2}}\right]\frac{1}{\theta
r}
\\
&~~~~-\frac{\lambda^{3}}{\theta^{3}}Z(\lambda
r)-\frac{\lambda\beta^{2}}{2\theta^{3}}Z_{2}(\lambda r),
\end{split}
\end{equation}
where
\[
\beta^{2}=\Lambda^{2}-m^{2},~~u^{2}=m^{2}-Q_{0}^{2},~~
\]
\[
\theta^{2}=-(m^{2}-Q_{0}^{2}),~~\lambda^{2}=\Lambda^{2}-Q_{0}^{2},
\]
and
\[
Y(x)=\frac{e^{-x}}{x},~~Z(x)=(1+\frac{3}{x}+\frac{3}{x^{2}})Y(x),
\]
\[
Z_{1}(x)=(\frac{1}{x}+\frac{1}{x^{2}})Y(x),~~Z_{2}(x)=(1+x)Y(x).
\]
The parameter $Q_0$ is the zero component of the four momentum of
exchanged meson.

We give some Fourier transformation formulae to derive the effective
potential,
\begin{equation}
\begin{split}
\frac{1}{u^{2}+\bm{Q}^{2}}\mathcal{F}^{2}(Q)&\rightarrow\frac{u}{4\pi}H_{0}(\Lambda,m,r),
\\
\frac{\bm{Q}^{2}}{u^{2}+\bm{Q}^{2}}\mathcal{F}^{2}(Q)&\rightarrow-\frac{u^{3}}{4\pi}H_{1}(\Lambda,m,r),
\\
\frac{\bm{Q}}{u^{2}+\bm{Q}^{2}}\mathcal{F}^{2}(Q)&\rightarrow\frac{iu^{3}}{4\pi}\boldsymbol{r}H_{2}(\Lambda,m,r),
\\
\frac{Q_{i}Q_{j}}{u^{2}+\bm{Q}^{2}}\mathcal{F}^{2}(Q)&\rightarrow-\frac{u^{3}}{12\pi}\left[H_{3}(\Lambda,m,r)K_{ij}+H_{1}(\Lambda,m,r)\delta_{ij}\right].
\end{split}
\end{equation}
If $u_{ex}^2=m_{ex}^2-Q_0^2<0$, the last formula above changes into
\[
\frac{Q_{i}Q_{j}}{u^{2}+\bm{Q}^{2}}\mathcal{F}^{2}(Q)\rightarrow-\frac{\theta^{3}}{12\pi}\left[M_{3}(\Lambda,m,r)K_{ij}+M_{1}(\Lambda,m,r)\delta_{ij}\right].
\]
%%%%%%%%%%%%%%%%%%%%%%%%%%%%%%
\section{Some Details of the Operators in the Lagrangian}\label{app_matrix}
%%%%%%%%%%%%%%%%%%%%%%%%%%%%%%%

We used some operators in the Eqs.~(\ref{V_S}-\ref{V_V}),
\begin{eqnarray}
\Delta_{S_AS_B}=\bm{\sigma}_{rsA}\cdot\bm{\sigma}_{rsB}, \quad
\Delta_{LS}={1 \over 2}\bm{L}\cdot\bm{\sigma}_{rs},\n
\\
\Delta_{T}={3\bm{\sigma}_{rsA}\cdot\bm{r}\bm{\sigma}_{rsB}\cdot\bm{r}
\over r^2}-\bm{\sigma}_{rsA}\cdot\bm{\sigma}_{rsB}.
\end{eqnarray}

$\Delta_{S_AS_B}$, $\Delta_{LS}$ and $\Delta_{T}$ are spin-spin
operator, spin-orbital operator and tensor operator, respectively.
$\bm{\sigma}_{rs}$ is the spin operator for spin-$\frac{3}{2}$
baryons. $\bm{L}$ is the relative orbit angular momentum operator
between the two baryons. $\bm{S}_A$ and $\bm{S}_B$ are the spin
operators of two baryons respectively, while
$\bm{S}=\bm{S}_A+\bm{S}_B$ is the total spin operator. For
spin-$\frac{3}{2}$ baryons, $\bm{S}=\frac{3}{2}\bm{\sigma}_{rs}$.

We introduce the transition spin operator $S^\mu_t$ for the
Rarita-Schwinger field $\Psi^\mu$, because we focus on the baryons
with spin $\frac{3}{2}$. The field $\Psi^\mu$ can be expressed as
\begin{equation}
  \Psi^\mu(\lambda)=\underset{m_\lambda}{\sum}\underset{m_s}{\sum}\epsilon^\mu(m_\lambda)\chi(m_s)=S^\mu_t\Phi,
\end{equation}
where $\epsilon^\mu(m_\lambda)$ is the polarization vector of a
spin-1 field,
\begin{equation}
\begin{split}
  \epsilon^\mu(+)&=-\frac{1}{\sqrt{2}}\left[0,1,i,0\right]^T,~
  \epsilon^\mu(0)=\left[0,0,0,1\right]^T,
  \\
  \epsilon^\mu(-)&=\frac{1}{\sqrt{2}}\left[0,1,-i,0\right]^T.
\end{split}
\end{equation}
$\chi$ is a two-component spinor. $\Phi$ is the spin wave function
of spin $\frac{3}{2}$ baryons.
\begin{equation}
\begin{split}
  \Phi\left(\frac{3}{2}\right)&=\left[1,0,0,0\right]^T,~
  \Phi\left(\frac{1}{2}\right)=\left[0,1,0,0\right]^T,
  \\
  \Phi\left(-\frac{1}{2}\right)&=\left[0,0,1,0\right]^T,~
  \Phi\left(-\frac{3}{2}\right)=\left[0,0,0,1\right]^T.
\end{split}
\end{equation}
It is easy to obtain the transition spin operator,
\begin{equation}
\begin{split}
  S^0_t=&0,~
  S^x_t=\frac{1}{\sqrt{2}}\left[
    \begin{array}{cccc}
    -1 & 0 & \frac{1}{\sqrt{2}} & 0\\
    0 & -\frac{1}{\sqrt{3}} & 0 & 1
    \end{array}\right],
  \\
  S^y_t=&-\frac{i}{\sqrt{2}}\left[
    \begin{array}{cccc}
    1 & 0 & \frac{1}{\sqrt{3}} & 0\\
    0 & \frac{1}{\sqrt{3}} & 0 & 1
    \end{array}\right],~
  S^x_t=\left[
    \begin{array}{cccc}
    0 & \sqrt{\frac{2}{3}} & 0 & 0\\
    0 & 0 & \sqrt{\frac{2}{3}} & 0
    \end{array}\right].
\end{split}
\end{equation}

The spin operator for spin-$3\over 2$ particles can be derived from
the Pauli matrices
$\bm{\sigma}_{rs}\equiv-S^\dagger_{t\mu}\bm{\sigma}S^\mu_t$. The
explicit form is
\begin{equation}
\begin{split}
  \sigma_{rs}^x&=\left[
    \begin{array}{cccc}
      0 & \frac{1}{\sqrt{3}} & 0 & 0 \\
      \frac{1}{\sqrt{3}} & 0 & \frac{2}{3} & 0 \\
      0 & \frac{2}{3} & 0 & \frac{1}{\sqrt{3}} \\
      0 & 0 & \frac{1}{\sqrt{3}} & 0
    \end{array}\right],~
  \sigma_{rs}^y=\left[
    \begin{array}{cccc}
      0 & -\frac{i}{\sqrt{3}} & 0 & 0 \\
      \frac{i}{\sqrt{3}} & 0 & -\frac{2i}{3} & 0 \\
      0 & \frac{2i}{3} & 0 & -\frac{i}{\sqrt{3}} \\
      0 & 0 & \frac{i}{\sqrt{3}} & 0
    \end{array}\right],
  \\
  \sigma_{rs}^z&=\left[
    \begin{array}{cccc}
      1 & 0 & 0 & 0 \\
      0 & \frac{1}{3} & 0 & 0 \\
      0 & 0 & -\frac{1}{3} & 0 \\
      0 & 0 & 0 & -1
    \end{array}\right]
\end{split}
\end{equation}
The tensor operator $\Delta_T$ is actually a scalar product of two
rank-2 tensor operators, $Y_{2,m}(\hat{r})$ and $T_{2,m}$
\begin{equation}
  \Delta_T=\overset{2}{\underset{m=-2}{\sum}}4\sqrt{\frac{6\pi}{5}}T_{2,m}Y_{2,m}^*(\hat{r})
\end{equation}
The operator $Y_{2,m}(\hat{r})$ is the spherical harmonic function,
and $T_{2,m}$ is a rank-2 tensor operator constructed by spin operator
\begin{equation}\label{2rank_tensor}
    \begin{split}
      &T_{2,\pm2}  =\frac{3}{8\pi}(S_x\pm iS_y)^2, \\
      &T_{2,\pm1}  =\mp\frac{3}{8\pi}[S_z(S_x\pm iS_y)+(S_x\pm iS_y)S_z], \\
      &T_{2,0}     =\frac{3}{4\sqrt{6}\pi}(3S_z^2-\bm{S}^2).
    \end{split}
\end{equation}
We can get the expression of the matrix elements of the tensor
operator
\begin{equation}
    \begin{split}
        &\braket{L_fS_fJ_fm_f|\Delta_T|L_iS_iJ_im_i}
        \\
        =&\underset{m_1}{\sum}\underset{m_2}{\sum}\underset{m_3}{\sum}\underset{m_4}{\sum}\braket {L_f\left[m_f-(m_3+m_4)\right],S_f(m_3+m_4)|J_fm_f}
        \\
        &\times\braket{ L_i\left[m_i-(m_1+m_2)\right],S_f(m_3+m_4)}
        \\
        &\times\braket{\frac{3}{2}m_3,\frac{3}{2}m_4|S_f(m_3+m_4)}\braket{\frac{3}{2}m_1,\frac{3}{2}m_2|S_i(m_1+m_2)}
        \\
        &\times\int^{2\pi}_0d\phi\int_{0}^{\pi}\sin(\theta)d\theta Y^*_{L_f,m_f-(m_3+m_4)}Y_{L_i,m_i-(m_1+m_2)}
        \\
        &\times\bigg[3\braket{\frac{3}{2}m_3|\vec{\sigma}_{rs1}|\frac{3}{2}m_1}\cdot\hat{r}(\theta,\phi)\braket{
        \frac{3}{2}m_4|\vec{\sigma}_{rs2}|\frac{3}{2}m_2}\cdot\hat{r}(\theta,\phi)
        \\
        &-\braket{\frac{3}{2}m_3|\vec{\sigma}_{rs1}|\frac{3}{2}m_1}\cdot
        \braket{\frac{3}{2}m_4|\vec{\sigma}_{rs2}|\frac{3}{2}m_2}\bigg].
    \end{split}
\end{equation}
The matrix elements of the tensor operator is independent of $m_i$
and $m_f$ according to the Wigner-Eckart theorem.

\section{Numerical Results of Baryon-antibaryon Systems}\label{app_BBbarsc}

We calculate the possible molecular states formed by one baryon and one antibaryon. In this section we perform the single channel calculation for baryon-antibaryon systems first. And then we take the multichannel effects for $J=0$ and $J=1$ systems. In this section, we only take the one-boson-exchange interaction into consideration. In fact, the three meson threshold may have a significant influence on the baryon-antibaryon systems. The threshold may change the existence or properties of the possible molecular states we obtained. Some of the binding solutions we obtained for the baryon-antibaryon systems may be narrow molecule-type resonances like X(3872).

\subsection{Single Channel Calculation}\label{subsubsec_BBbar_sc}

\begin{figure*}[htp]
\centering
\includegraphics[width=0.95\textwidth]{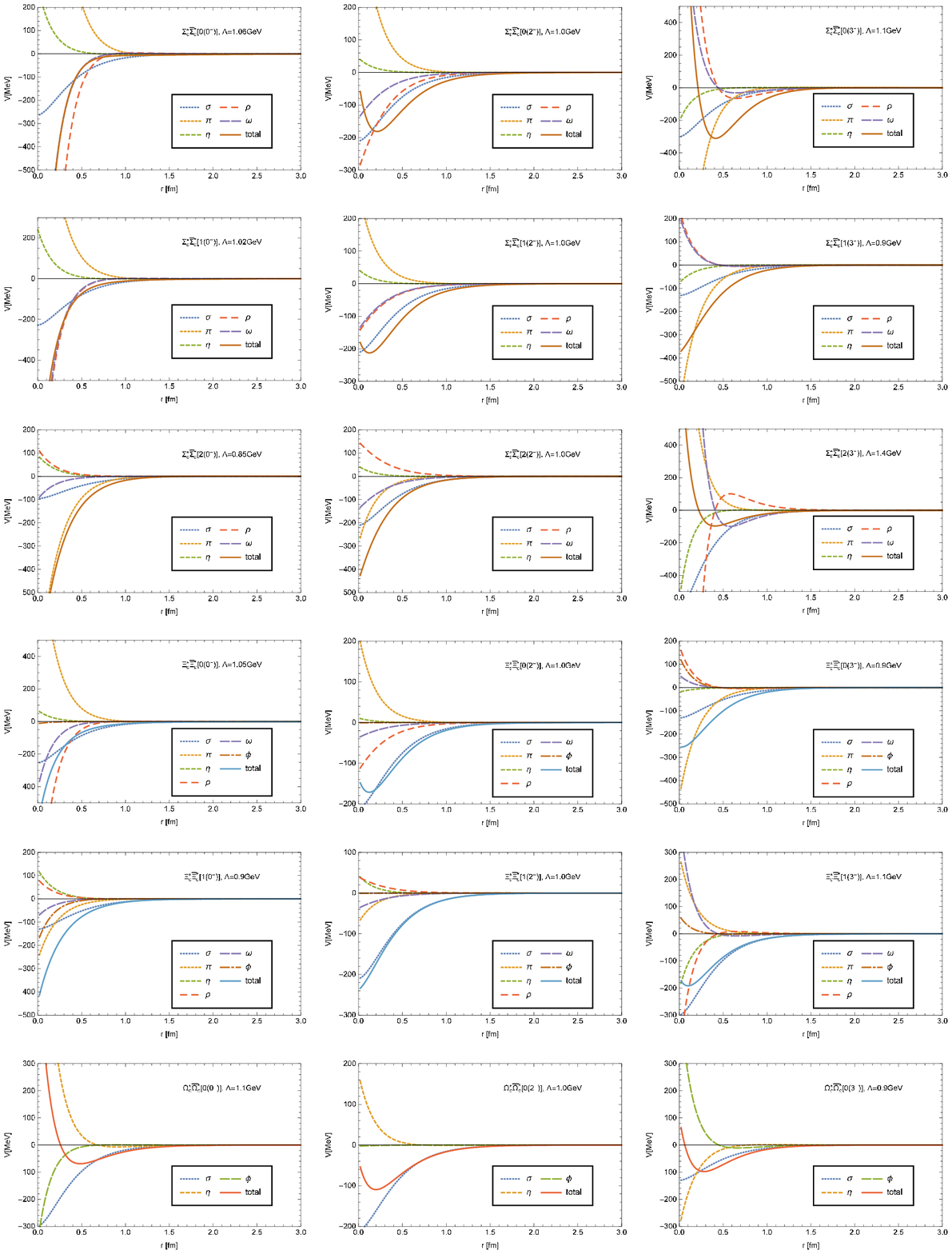}
\caption{The interaction potentials for the baryon-antibaryon
single channel systems.}\label{fig_p_BBbar_sc}
\end{figure*}

The numerical results of the baryon-antibaryon systems are collected in Table~\ref{Table_num_BBbar_sc}. The relevant potentials are shown in Fig.~\ref{fig_p_BBbar_sc}. We find some candidates of molecular states when the cutoff parameters are suitable.

\begin{table*}[htp]
\centering \caption{The numerical results for baryon-antibaryon
single channel systems. $\Lambda $ is the cutoff parameter. "$E$" is
the binding energy. $R_{rms}$ is the root-mean-square radius. We use
$[I(J^P)]$ to mark different states.}\label{Table_num_BBbar_sc}
\begin{tabular}{cccc|cccc}
\hline \hline
States & $\Lambda$(MeV) & E(MeV) & $R_{rms}$(fm) & States & $\Lambda$(MeV) & E(MeV) & $R_{rms}$(fm)\tabularnewline
\hline
\multirow{3}{*}{$\Sigma_{c}^{*}\bar{\Sigma}_{c}^{*}[0(0^{-})]$} & \multirow{1}{*}{1040} & 13.9 & 1.0 & \multirow{3}{*}{$\Xi_{c}^{*}\bar{\Xi}_{c}^{*}[0(0^{-})]$} & 1000 & 0.4 & 4.4\tabularnewline
 & 1060 & 76.0 & 0.5 &  & 1050 & 10.4 & 1.2\tabularnewline
 & 1080 & 185.2 & 0.4 &  & 1100 & 56.3 & 0.6\tabularnewline
\hline
\multirow{3}{*}{$\Sigma_{c}^{*}\bar{\Sigma}_{c}^{*}[0(1^{-})]$} & 1020 & 5.7 & 1.6 & \multirow{3}{*}{$\Xi_{c}^{*}\bar{\Xi}_{c}^{*}[0(1^{-})]$} & 1000 & 2.5 & 2.2\tabularnewline
 & 1040 & 23.6 & 0.9 &  & 1050 & 15.2 & 1.0\tabularnewline
 & 1060 & 62.2 & 0.6 &  & 1100 & 49.9 & 0.6\tabularnewline
\hline
\multirow{3}{*}{$\Sigma_{c}^{*}\bar{\Sigma}_{c}^{*}[0(2^{-})]$} & 950 & 6.6 & 1.5 & \multirow{3}{*}{$\Xi_{c}^{*}\bar{\Xi}_{c}^{*}[0(2^{-})]$} & 950 & 4.5 & 1.7\tabularnewline
 & 1000 & 25.6 & 0.9 &  & 1000 & 12.8 & 1.2\tabularnewline
 & 1050 & 58.3 & 0.7 &  & 1050 & 26.4 & 0.9\tabularnewline
\hline
\multirow{3}{*}{$\Sigma_{c}^{*}\bar{\Sigma}_{c}^{*}[0(3^{-})]$} & 1000 & 103.6 & 0.6 & \multirow{3}{*}{$\Xi_{c}^{*}\bar{\Xi}_{c}^{*}[0(3^{-})]$} & 800 & 4.7 & 1.6\tabularnewline
 & 1100 & 111.8 & 0.7 &  & 900 & 21.1 & 0.9\tabularnewline
 & 1200 & 139.6 & 0.7 &  & 1000 & 39.2 & 0.8\tabularnewline
\hline
\multirow{3}{*}{$\Sigma_{c}^{*}\bar{\Sigma}_{c}^{*}[1(0^{-})]$} & 1000 & 6.0 & 1.5 & \multirow{3}{*}{$\Xi_{c}^{*}\bar{\Xi}_{c}^{*}[1(0^{-})]$} & 800 & 23.0 & 0.8\tabularnewline
 & 1020 & 24.1 & 0.8 &  & 900 & 11.1 & 1.1\tabularnewline
 & 1040 & 60.4 & 0.6 &  & 1000 & 18.6 & 1.0\tabularnewline
\hline
\multirow{3}{*}{$\Sigma_{c}^{*}\bar{\Sigma}_{c}^{*}[1(1^{-})]$} & 980 & 3.4 & 2.0 & \multirow{3}{*}{$\Xi_{c}^{*}\bar{\Xi}_{c}^{*}[1(1^{-})]$} & 800 & 10.3 & 1.1\tabularnewline
 & 1000 & 10.8 & 1.2 &  & 900 & 7.1 & 1.4\tabularnewline
 & 1020 & 25.4 & 0.9 &  & 1000 & 15.1 & 1.0\tabularnewline
\hline
\multirow{3}{*}{$\Sigma_{c}^{*}\bar{\Sigma}_{c}^{*}[1(2^{-})]$} & 950 & 7.2 & 1.5 & \multirow{3}{*}{$\Xi_{c}^{*}\bar{\Xi}_{c}^{*}[1(2^{-})]$} & 900 & 2.1 & 2.3\tabularnewline
 & 1000 & 22.9 & 1.0 &  & 1000 & 9.2 & 1.3\tabularnewline
 & 1050 & 47.9 & 0.7 &  & 1100 & 24.3 & 0.9\tabularnewline
\hline
\multirow{3}{*}{$\Sigma_{c}^{*}\bar{\Sigma}_{c}^{*}[1(3^{-})]$} & 800 & 13.9 & 1.0 & \multirow{3}{*}{$\Xi_{c}^{*}\bar{\Xi}_{c}^{*}[1(3^{-})]$} & 1000 & 3.2 & 1.9\tabularnewline
 & 900 & 33.6 & 0.8 &  & 1100 & 12.3 & 1.1\tabularnewline
 & 1000 & 45.3 & 0.8 &  & 1200 & 23.6 & 0.9\tabularnewline
\hline
\multirow{3}{*}{$\Sigma_{c}^{*}\bar{\Sigma}_{c}^{*}[2(0^{-})]$} & 800 & 41.1 & 0.7 & \multirow{3}{*}{$\Omega_{c}^{*}\bar{\Omega}_{c}^{*}[0(0^{-})]$} & 1000 & 0.8 & 3.5\tabularnewline
 & 850 & 62.4 & 0.6 &  & 1100 & 4.6 & 1.8\tabularnewline
 & 900 & 88.3 & 0.5 &  & 1200 & 25.5 & 0.9\tabularnewline
\hline
\multirow{3}{*}{$\Sigma_{c}^{*}\bar{\Sigma}_{c}^{*}[2(1^{-})]$} & 800 & 18.0 & 1.0 & \multirow{3}{*}{$\Omega_{c}^{*}\bar{\Omega}_{c}^{*}[0(1^{-})]$} & 1000 & 1.4 & 2.8\tabularnewline
 & 850 & 31.6 & 0.8 &  & 1100 & 6.8 & 1.5\tabularnewline
 & 900 & 49.0 & 0.7 &  & 1200 & 31.7 & 0.8\tabularnewline
\hline
\multirow{3}{*}{$\Sigma_{c}^{*}\bar{\Sigma}_{c}^{*}[2(2^{-})]$} & 900 & 4.4 & 1.7 & \multirow{3}{*}{$\Omega_{c}^{*}\bar{\Omega}_{c}^{*}[0(2^{-})]$} & 900 & 4.2 & 1.7\tabularnewline
 & 1000 & 18.2 & 1.0 &  & 1000 & 5.3 & 1.6\tabularnewline
 & 1100 & 39.6 & 0.7 &  & 1100 & 17.4 & 1.0\tabularnewline
\hline
\multirow{3}{*}{$\Sigma_{c}^{*}\bar{\Sigma}_{c}^{*}[2(3^{-})]$} & 1300 & 2.5 & 2.3 & \multirow{3}{*}{$\Omega_{c}^{*}\bar{\Omega}_{c}^{*}[0(3^{-})]$} & 800 & 4.1 & 1.9\tabularnewline
 & 1400 & 5.4 & 1.7 &  & 900 & 4.9 & 1.6\tabularnewline
 & 1500 & 9.1 & 1.4 &  & 1000 & 34.2 & 0.7\tabularnewline
\hline \hline
\end{tabular}

\end{table*}

For the $\Sigma_{c}^{*}\bar{\Sigma}_{c}^{*}$ system, the potential arises from the $\sigma$, $\pi$, $\eta$, $\rho$ and $\omega$ exchanges. They are all candidates of molecular states when some suitable cutoff parameters are chosen. We find binding solutions for the $\Sigma_{c}^{*}\bar{\Sigma}_{c}^{*}[0(0^-)$, $0(1^-)$, $1(0^-)]$ systems, when the cutoff parameters are around 1.0 GeV. The solutions are more dependent on the cutoff than other systems. For the $\Sigma_{c}^{*}\bar{\Sigma}_{c}^{*}[0(3^-)]$ system, we find a solution with binding energy 103.6-139.6 MeV, when the cutoff parameter varies from 1.0 GeV to 1.2 GeV. The binding energy is less dependent on the cutoff. For the $\Sigma_{c}^{*}\bar{\Sigma}_{c}^{*}[1(2^-)]$ system, the $\sigma$, $\rho$ and $\omega$ exchange potentials are attractive, while the $\pi$, $\eta$ exchange potentials are weakly repulsive. We find a molecular solution with the binding energy 7.2-47.9 MeV when the cutoff parameter is from 0.95 GeV to 1.05 GeV. For the $\Sigma_{c}^{*}\bar{\Sigma}_{c}^{*}[2(0^-)]$ system, the $\pi$ exchange is dominant, which is repulsive in the range $r<1$ fm. The binding energy is 41.1-88.3 MeV when the cutoff parameter is 0.8-0.9 GeV. For the $\Sigma_{c}^{*}\bar{\Sigma}_{c}^{*}[2(3^-)]$ system, the contributions of the $\rho$ and $\omega$ exchanges cancel out. The contributions from other mesons make the total potential slightly attractive. As a result, we find a binding solution with a small binding energy, 2.5-9.1 MeV, when the cutoff parameter is 1.3-1.5 GeV.

For the $\Xi_{c}^{*}\bar{\Xi}_{c}^{*}$ systems, we get some loosely binding solutions when the cutoff is about 1 GeV. They are candidates of molecular states. Compared with the $\Sigma_{c}^{*}\bar{\Sigma}_{c}^{*}$ systems, the $\Xi_{c}^{*}\bar{\Xi}_{c}^{*}$ system also allows the $\phi$ meson exchange, although it's contribution is usually small. For the $\Xi_{c}^{*}\bar{\Xi}_{c}^{*}[0(0^-)$, $0(1^-)$, $0(2^-)]$ systems, the $\pi$ exchange provides the repulsive part of the total potential. The attractive part mainly arises from the $\sigma$, $\rho$ and $\omega$ exchanges. Their binding energies and $R_{rms}$ are shown in Table\ref{Table_num_BBbar_sc}. For the $\Xi_{c}^{*}\bar{\Xi}_{c}^{*}[0(1^-)]$ system, the $\pi$ and $\sigma$ exchange provide the weakly attractive potential. The binding energy is 4.7-39.2 MeV with the cutoff parameter from 0.8 GeV to 1.0 GeV. For the $\Xi_{c}^{*}\bar{\Xi}_{c}^{*}[1(2^-)]$ system, the attractive potential mainly comes from the $\sigma$ exchange. A molecular solution with the binding energy 2.1-24.3 MeV appears when the cutoff is 0.9-1.1 GeV. For the $\Xi_{c}^{*}\bar{\Xi}_{c}^{*}[1(3^-)]$ system. We get the numerical result with the binding energy 3.2-23.6 MeV, when the cutoff parameter varies form 1.0 GeV to 1.2 GeV.

For the $\Omega_{c}^{*}\bar{\Omega}_{c}^{*}$ systems with spin 0,1,2, we get some deuteron-like solutions. All the $\Omega_{c}^{*}\bar{\Omega}_{c}^{*}$ systems are expected to be candidates of molecular states. From the Fig.~\ref{fig_p_BBbar_sc}, it seems that $\sigma$ exchange provides most of the potential. The total potential is weakly attractive in the medium and long range. The binding energy of
$\Omega_{c}^{*}\bar{\Omega}_{c}^{*}[0(0^{-})]$ is 0.8-25.5 MeV when
the cutoff is from 1.0 GeV to 1.2 GeV. The binding energy of
$\Omega_{c}^{*}\bar{\Omega}_{c}^{*}[0(1^{-})]$ is 1.4-31.7 MeV when
the cutoff parameter is in the range of 1.0 GeV-1.2 GeV. For the
$\Omega_{c}^{*}\bar{\Omega}_{c}^{*}[0(2^{-})]$, the binding energy
is 4.2-17.4 MeV when the cutoff parameter changes from 0.9 GeV to 1.1 GeV.
For the $\Omega_{c}^{*}\bar{\Omega}_{c}^{*}[0(3^{-})]$, the main
part of the potential comes from the $\sigma$ and $\eta$ exchanges. The binding
energy is 4.1-34.2 MeV when the cutoff is from 0.8 GeV to 1.0
GeV.

\subsection{Couple Channel Calculation}\label{subsubsec_BBbar_cc}

\begin{figure*}[htp]
\centering
\includegraphics[width=0.95\textwidth]{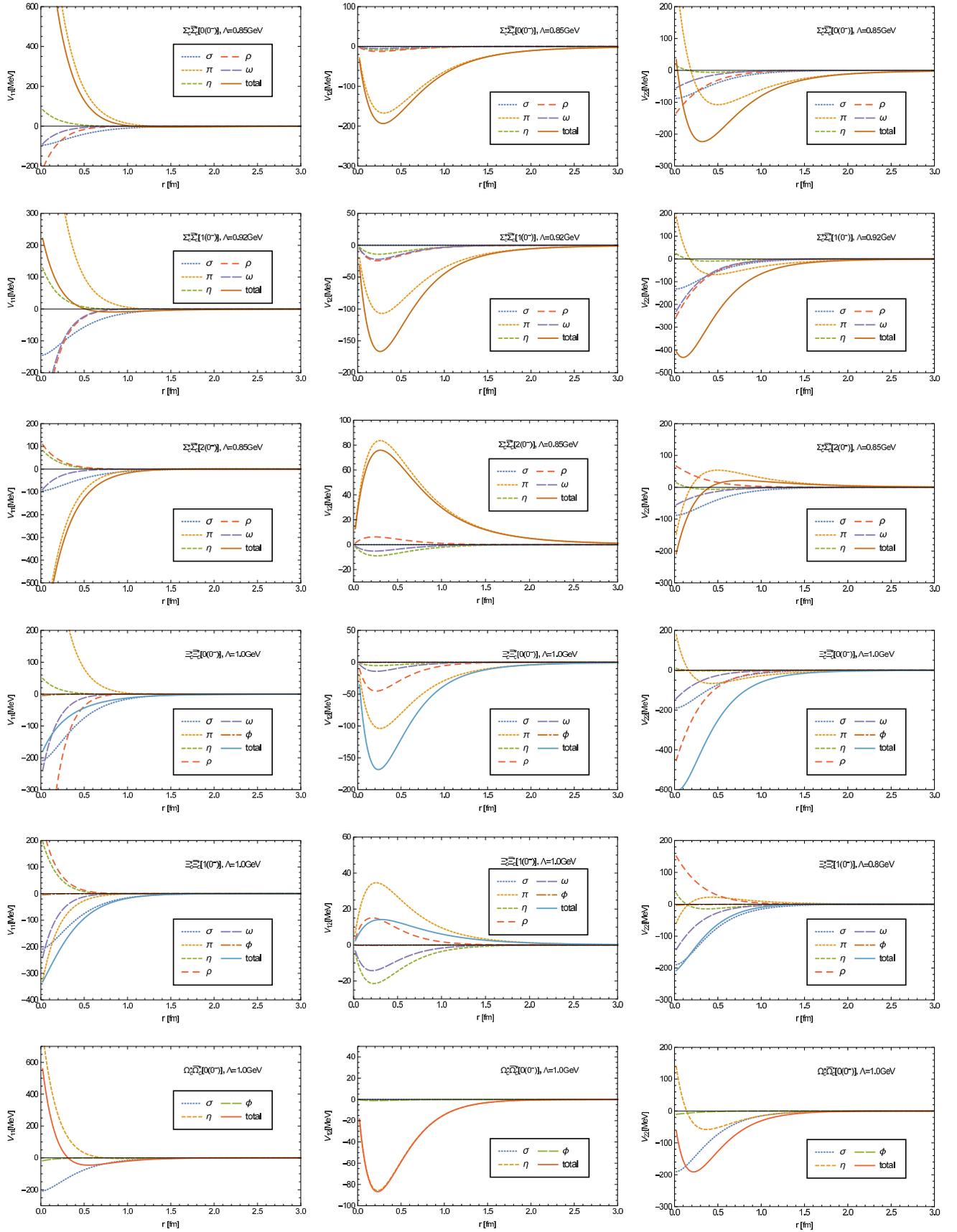}
\caption{Some typical interaction potentials for the
baryon-antibaryon couple channel systems with total spin 0.
$V_{11},V_{12}$ and $V_{22}$ denote the
${}^{1}S_{0}\leftrightarrow{}^{1}S_{0},{}^{1}S_{0}\leftrightarrow{}^{5}D_{0}$
and ${}^{5}D_{0}\leftrightarrow{}^{5}D_{0}$ transitions potentials.}\label{fig_p_BBbar_cc1}
\end{figure*}

\begin{figure*}[htp]
\centering
\includegraphics[width=0.95\textwidth]{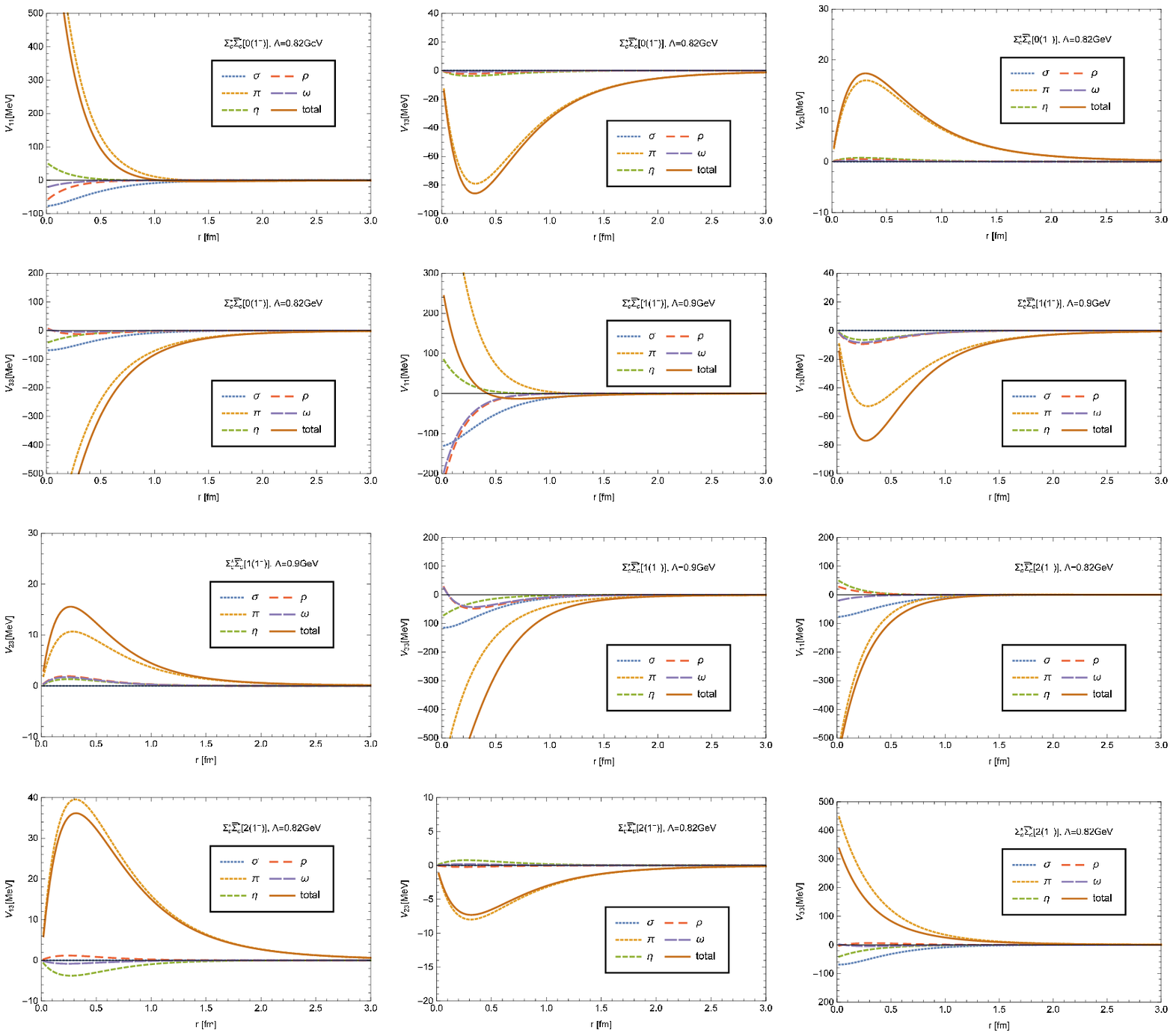}
\caption{Some typical interaction potentials for the
$\Sigma_{c}^{*}\bar{\Sigma}_{c}^{*}$ couple channel systems with
total spin 1. The subscript number 1-4 means
states${}^3S_{1},{}^3D_{1},{}^7D_{1}$ and ${}^7G_{1}$ in sequence.
We put only four representative potentials here. For other
potentials, $V_{12}$ and $V_{22}$ are similar to $V_{11}$, $V_{24}$
is similar to $V_{13}$, while $V_{34}$ and $V_{44}$ are similar to
$V_{34}$.}\label{fig_p_BBbar_cc2}
\end{figure*}

\begin{figure*}[htp]
\centering
\includegraphics[width=0.95\textwidth]{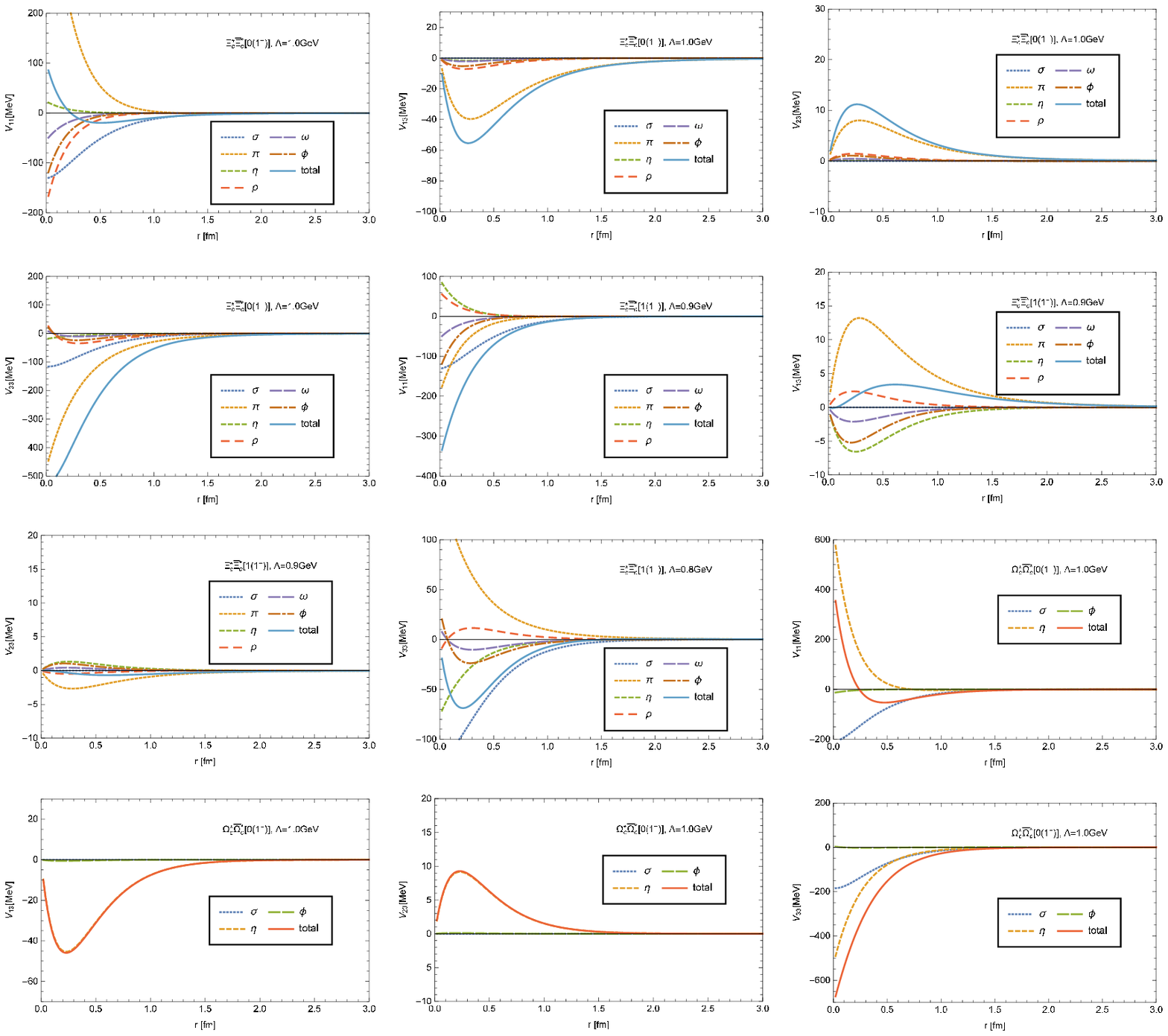}
\caption{Some typical interaction potentials for the
$\Xi_{c}^{*}\bar{\Xi}_{c}^{*}$ and
$\Omega_{c}^{*}\bar{\Omega}_{c}^{*}$ couple channel systems with
total spin 1. The subscript number 1-4 means
states${}^3S_{1},{}^3D_{1},{}^7D_{1}$ and ${}^7G_{1}$ in sequence.
We put only four representative potentials here. For other
potentials, $V_{12}$ and $V_{22}$ are similar to $V_{11}$, $V_{24}$
is similar to $V_{13}$, while $V_{34}$ and $V_{44}$ are similar to
$V_{34}$.}\label{fig_p_BBbar_cc3}
\end{figure*}

In this subsection, the couple channel effect is added for $J=0,1$
systems. The numerical results are shown in
Tables~\ref{Table_num_BBbar_cc1} and \ref{Table_num_BBbar_cc2}
respectively. The corresponding potentials are given in
Figs.~\ref{fig_p_BBbar_cc1}-\ref{fig_p_BBbar_cc3}.

\begin{table*}[htp]
\centering \caption{The numerical results for two baryon-antibaryon
couple channel systems with total spin 0. $\Lambda $ is the cutoff
parameter. "$E$" is the binding energy. $R_{rms}$ is the
root-mean-square radius. We use $[I(J^P)]$ to mark different states.
$P_{S}$ is the percentage of the S wave, and $P_{D}$ is the
percentage of the D wave.}\label{Table_num_BBbar_cc1}
\begin{tabular}{cccccc|cccccc}
\hline \hline
States & $\Lambda$(MeV) & E(MeV) & $R_{rms}$(fm) & $P_{S}(\%)$ & $P_{D}(\%)$ & States & $\Lambda$(MeV) & E(MeV) & $R_{rms}$(fm) & $P_{S}(\%)$ & $P_{D}(\%)$\tabularnewline
\hline
\multirow{3}{*}{$\Sigma_{c}^{*}\bar{\Sigma}_{c}^{*}[0(0^{-})]$} & 800 & 5.1 & 3.0 & 71.7 & 28.3 & \multirow{3}{*}{$\Xi_{c}^{*}\bar{\Xi}_{c}^{*}[0(0^{-})]$} & 950 & 5.5 & 2.2 & 86.8 & 13.2\tabularnewline
 & 850 & 14.9 & 2.1 & 60.9 & 39.1 &  & 1000 & 21.4 & 1.4 & 80.5 & 19.5\tabularnewline
 & 900 & 43.8 & 1.5 & 50.6 & 49.4 &  & 1050 & 61.1 & 1.0 & 77.4 & 22.6\tabularnewline
\hline
\multirow{3}{*}{$\Sigma_{c}^{*}\bar{\Sigma}_{c}^{*}[1(0^{-})]$} & 900 & 4.2 & 2.6 & 83.1 & 16.9 & \multirow{3}{*}{$\Xi_{c}^{*}\bar{\Xi}_{c}^{*}[1(0^{-})]$} & 900 & 11.2 & 1.2 & 99.9 & 0.1\tabularnewline
 & 920 & 9.9 & 2.0 & 77.3 & 22.7 &  & 1000 & 18.9 & 1.0 & 99.8 & 0.2\tabularnewline
 & 940 & 20.1 & 1.6 & 72.5 & 27.5 &  & 1100 & 40.9 & 0.8 & 99.9 & 0.1\tabularnewline
\hline
\multirow{3}{*}{$\Sigma_{c}^{*}\bar{\Sigma}_{c}^{*}[2(0^{-})]$} & 800 & 45.6 & 0.8 & 98.7 & 1.3 & \multirow{3}{*}{$\Omega_{c}^{*}\bar{\Omega}_{c}^{*}[0(0^{-})]$} & 900 & 7.8 & 1.5 & 97.1 & 2.9\tabularnewline
 & 850 & 67.9 & 0.7 & 98.7 & 1.3 &  & 1000 & 2.3 & 2.4 & 98.3 & 1.7\tabularnewline
 & 900 & 94.9 & 0.6 & 98.8 & 1.2 &  & 1100 & 13.2 & 1.4 & 93.5 & 6.5\tabularnewline
\hline \hline
\end{tabular}
\end{table*}

\begin{table*}[htp]
\centering \caption{The numerical results for two baryon-antibaryon
couple channel systems with total spin 1. $\Lambda $ is the cutoff
parameter. "$E$" is the binding energy. $R_{rms}$ is the
root-mean-square radius. We use $[I(J^P)]$ to mark different states.
$P_{S}$ is the percentage of the ${}^3S_1$, $P_{D1}$ is the
percentage of the ${}^3D_1$, $P_{D2}$ is the percentage of the
${}^7D_1$, and $P_{G}$ is the percentage of the
${}^7G_1$.}\label{Table_num_BBbar_cc2}
\begin{tabular}{cccccccc}
\hline \hline
States & $\Lambda$(MeV) & E(MeV) & $R_{rms}$(fm) & $P_{S}(\%)$ & $P_{D1}(\%)$ & $P_{D2}(\%)$ & $P_{G}(\%)$\tabularnewline
\hline
\multirow{3}{*}{$\Sigma_{c}^{*}\bar{\Sigma}_{c}^{*}[0(1^{-})]$} & 800 & 5.1 & 3.3 & 72.0 & 13.1 & 14.8 & 0.1\tabularnewline
 & 820 & 8.2 & 2.9 & 67.1 & 14.0 & 18.7 & 0.2\tabularnewline
 & 840 & 13.5 & 2.6 & 61.6 & 14.5 & 23.7 & 0.2\tabularnewline
\hline
\multirow{3}{*}{$\Sigma_{c}^{*}\bar{\Sigma}_{c}^{*}[1(1^{-})]$} & 880 & 1.9 & 3.5 & 89.2 & 5.7 & 5.1 & 0.0\tabularnewline
 & 900 & 5.4 & 2.6 & 83.8 & 7.8 & 8.3 & 0.1\tabularnewline
 & 940 & 22.8 & 1.8 & 74.2 & 10.1 & 15.6 & 0.1\tabularnewline
\hline
\multirow{3}{*}{$\Sigma_{c}^{*}\bar{\Sigma}_{c}^{*}[2(1^{-})]$} & 800 & 22.6 & 1.1 & 98.0 & 1.5 & 0.5 & 0.0\tabularnewline
 & 820 & 28.1 & 1.0 & 98.1 & 1.5 & 0.4 & 0.0\tabularnewline
 & 840 & 34.2 & 1.0 & 98.1 & 1.5 & 0.4 & 0.0\tabularnewline
\hline
\multirow{3}{*}{$\Xi_{c}^{*'}\bar{\Xi}_{c}^{*'}[0(1^{-})]$} & 900 & 1.5 & 3.5 & 93.4 & 3.7 & 2.9 & 0.0\tabularnewline
 & 950 & 7.5 & 2.1 & 88.0 & 6.1 & 5.9 & 0.0\tabularnewline
 & 1000 & 24.2 & 1.5 & 82.9 & 7.7 & 9.4 & 0.0\tabularnewline
\hline
\multirow{3}{*}{$\Xi_{c}^{*'}\bar{\Xi}_{c}^{*'}[1(1^{-})]$} & 800 & 10.4 & 1.2 & 100.0 & 0.0 & 0.0 & 0.0\tabularnewline
 & 900 & 7.2 & 1.4 & 99.9 & 0.1 & 0.0 & 0.0\tabularnewline
 & 1000 & 15.4 & 1.1 & 99.8 & 0.1 & 0.1 & 0.0\tabularnewline
\hline
\multirow{3}{*}{$\Omega_{c}^{*}\bar{\Omega}_{c}^{*}[0(1^{-})]$} & 900 & 7.4 & 1.6 & 97.5 & 1.5 & 1.0 & 0.0\tabularnewline
 & 1000 & 3.2 & 2.2 & 98.3 & 1.1 & 0.6 & 0.0\tabularnewline
 & 1100 & 15.8 & 1.4 & 94.2 & 3.5 & 2.3 & 0.0\tabularnewline
\hline \hline
\end{tabular}
\end{table*}

For the states with total spin 0, we consider the couple channel effect for the six systems, $\Sigma_{c}^{*}\bar{\Sigma}_{c}^{*}$ with isospin 0, 1, 2, and $\Xi_{c}^{*}\bar{\Xi}_{c}^{*}$ with isospin 0, 1 as well as $\Omega_{c}^{*}\bar{\Omega}_{c}^{*}[0(0^{-})]$. They are all candidates of molecular states. For most cases, the D-wave channel affects the S-wave channel slightly. For the $\Xi_{c}^{*}\bar{\Xi}_{c}^{*}[1(0^{-})]$ system, as shown in Table~\ref{Table_num_BBbar_cc1}, the D-wave contribution in the total wave function is about 0.1\%, and makes the binding energy shift 0.2 MeV when the cutoff parameter is 0.9 GeV. The $\Sigma_{c}^{*}\bar{\Sigma}_{c}^{*}$$[0(0^{-})]$ system is interesting. The couple channel effect does change the result quite a lot. The D-wave contribution is around 40\%. For the single channel case, we choose the cutoff parameter from 1.04 GeV to 1.06 GeV. After considering the couple channel effect, we choose a new range of cutoff parameter, 0.8-0.9 GeV, to get the binding solutions with reasonable small binding energies, 5.1-43.8 MeV. For the $\Sigma_{c}^{*}\bar{\Sigma}_{c}^{*}[1(0^{-})]$ system, we find a loosely bound solution, whose binding energy is 4.2-20.1 MeV, when the cutoff parameter is 0.9-0.94 GeV. The couple channel effect also has a significant influence on the system. For the $\Sigma_{c}^{*}\bar{\Sigma}_{c}^{*}[2(0^{-})]$ system, the couple channel effect only makes the binding a little deeper. For the $\Xi_{c}^{*}\bar{\Xi}_{c}^{*}[0(0^{-})]$ system, the binding energy is 5.5-61.1 MeV while the cutoff parameter is 0.95-1.05 MeV. For the $\Omega_{c}^{*}\bar{\Omega}_{c}^{*}[0(0^{-})]$ system, a molecular solution appears when the cutoff parameter varies from 0.9 GeV to 1.1 GeV.

For the system with spin 1, we add G-wave besides the S- and D-waves. All the six systems are candidates of molecular states. For the $\Sigma_{c}^{*}\bar{\Sigma}_{c}^{*}[0(1^{-})]$ system, the D-waves have a nontrivial influence. The contribution of D-waves is almost 40\%, when the cutoff is 0.84 GeV. The large D-waves contribution makes us choose the different cutoff parameters from the single channel case. The binding energy for multichannel calculation is 5.1-13.5 MeV, while the cutoff parameter is 0.8-0.84 GeV. For the $\Sigma_{c}^{*}\bar{\Sigma}_{c}^{*}[1(1^{-})]$ system, the effect of D-waves is also obvious. The binding energy is 1.9-22.8 MeV when the cutoff is 8.8-9.4 GeV. The $\Xi_{c}^{*}\bar{\Xi}_{c}^{*}[0(1^-)]$ system is similiar. We find that a loosely binding solution with binding energy 1.5-24.2 MeV appears when the cutoff parameter is 0.9-1.0 GeV. For the $\Sigma_{c}^{*}\bar{\Sigma}_{c}^{*}[1(1^{-})]$ system, the S-wave dominates the total wave function. The binding energy is larger than the single channel calculation. For the $\Xi_{c}^{*}\bar{\Xi}_{c}^{*}[1(1^-)]$ system, the D-wave contribution is less than 0.2\%, the binding energy is also larger than that in the single channel case. For the $\Omega_{c}^{*}\bar{\Omega}_{c}^{*}[0(1^{-})]$ state, the couple channel effect makes the binding deeper as expected. The binding energy is 3.2 MeV while the cutoff is 1.0 GeV.

\subsection{Summary}

We calculate the baryon-antibaryon systems with different spin and isospin in single channel, and find the loosely bound solutions. After considering the channel mixing effect, we calculate the systems with total spin 0 and 1. For the most systems, the multichannel effect would lead to a deeper binding solution. For the $\Sigma_{c}^{*}\bar{\Sigma}_{c}^{*}$$[0(0^{-})$, $1(0^{-})$, $0(1^{-})$, $1(1^{-})]$ and $\Xi_{c}^{*}\Xi_{c}^{*}$$[0(1-)]$ systems, the D-wave contribution is nontrivial, and may even reaches up to 40\%.

Moreover, a baryon-antibaryon molecular state may also decay into three mesons through quark rearrangement, which makes the molecular states unstable. Some of the ``bound sates'' obtained in this section may appear as other structures in experiment considering the open three mesons threshold. On the one hand, some of these binding solutions may appear as a possible enhancement of the baryon and antibaryon invariant mass spectrum in experiment, instead of as a real resonance. On the other hand, some of these binding solutions may appear as a narrow resonance state like X(3872). X(3872) is a good candidate of the $D\bar{D}^*$ molecule. Although it decays into $D\bar{D}\pi$, X(3872) is still a very narrow resonance. Another example, the charged $Z_c$ states containing four quarks, are above some two mesons thresholds. Even though the $Z_c$ states decay into two mesons through quark rearrangement, they still appear as rather narrow resonances in experiment.
\end{appendix}

\end{document}